\documentclass[preprint, 3p, times, 12pt]{elsarticle}
\usepackage{amssymb}
\usepackage{amsmath}
\usepackage{tabularx}
\usepackage{booktabs}
\usepackage{caption}
\usepackage[colorlinks=true, linkcolor=blue, citecolor=blue, urlcolor=blue]{hyperref}
\journal{Journal of Computational Physics}

\begin{document}
\begin{frontmatter}
\title{Direct numerical simulation of two-phase flows with surfactant-induced surface viscous effects}
\author[label1]{Debashis Panda}
\author[label2]{Seungwon Shin$^*$}
\author[label1,label3]{Abdullah M. Abdal}
\author[label1]{Lyes Kahouadji}
\author[label4]{Jalel Chergui}
\author[label4,label5]{Damir Juric$^{**}$}
\author[label1]{Omar K. Matar}

\affiliation[label1]{organization={Department of Chemical Engineering, Imperial College London, South Kensington campus},
            addressline={}, 
            city={London},
            postcode={SW7 2AZ}, 
            country={United Kingdom}}

\affiliation[label2]{organization={Department of Mechanical and System Design Engineering, Hongik University},
            addressline={72-1, Sangsu-dong, Mapo-gu}, 
            city={Seoul},
            postcode={127-791}, 
            country={Republic of Korea}}
\affiliation[label3]{organization={Department of Environmental and Sustainability Engineering, College of Engineering and Energy, Abdullah Al Salem University},
            city={Kuwait City},
            postcode={12037},  
            country={Kuwait}}

\affiliation[label4]{organization={Université Paris Saclay, Centre National de la Recherche Scientiéfique (CNRS), Laboratoire Interdisciplinaire des Sciences du Numerique (LISN)},
            city={Orsay},
            postcode={91400}, 
            country={France}} 
\affiliation[label5]{organization={Department of Applied Mathematics and Theoretical Physics and Centre for Mathematical Sciences, University of Cambridge},
            addressline={Wilberforce Road}, 
            city={Cambridge},
            postcode={CB3 0WA}, 
            country={United Kingdom}} 
\cortext[cor1]{Corresponding author: Seungwon Shin (sshin@hongik.ac.kr)}
\cortext[cor2]{Corresponding author: Damir Juric (damir.juric@lisn.fr)}

\begin{abstract}
Direct numerical simulations of interfacial flows with surfactant-induced complexities involving surface viscous stresses are performed within the framework of the Level Contour Reconstruction Method (LCRM); this hybrid front-tracking/level-set approach leverages the advantages of both methods. In addition to interface-confined surfactant transport that results in surface diffusion and Marangoni stresses, the interface, endowed with shear and dilatational viscosities; these act to resist deformation arising from 
velocity gradients in the plane of the two-dimensional manifold of the interface, and interfacial compressibility effects, respectively. 
By adopting the Boussinesq-Scriven constitutive model, We provide a mathematical formulation of these effects that accurately captures the interfacial mechanics, which is then implemented within the LCRM-based code by exploiting the benefits inherent to the underlying front-tracking/level-sets hybrid approach. 
We validate our numerical predictions against a number of benchmark cases that involve drops undergoing deformation when subjected to a flow field or when rising under the action of buoyancy. The results of these validation studies highlight the importance of adopting a rigorous approach in modelling the interfacial dynamics. We also present results that demonstrate the effects of surface viscous stresses on interfacial deformation in unsteady parametric surface waves and atomisation events. 
%
\end{abstract}



\end{frontmatter}

\begin{table*}[!h]
\caption*{Nomenclature}
    \footnotesize
    \begin{tabular}{ccc ||ccc}
    \hline
    \textbf{Variables} & \textbf{Notation} & \textbf{Unit} & 
    \textbf{Variables} & \textbf{Notation} & \textbf{Unit} \\
    \hline
    \multicolumn{6}{c}{\textbf{Greek letters}} \\
    \hline
    Viscosity of two phases & $\mu_1, \mu_2$ & Pa.s &
    Density of two phases & $\rho_1, \rho_2$ & kg m$^{-3}$ \\
    Dilatational bulk viscosity & $\mu_d^\mathbf X$ & Pa.s &
    Shear bulk viscosity & $\mu_s^\mathbf X$ & Pa.s \\
    Dilatational surface viscosity & $\mu_d^{\mathbf S}$ & Pa.s.m &
    Shear surface viscosity & $\mu_s^{\mathbf S}$ & Pa.s.m \\
    One-fluid density & $\rho$ & kg m$^{-3}$ &
    One-fluid viscosity & $\mu$ & Pa.s \\
    Kronecker delta & $\delta$ & -- &
    Surfactant concentration & $\Gamma$ & mol.m$^{-2}$ \\
    Surface tension & $\sigma$ & N/m &
    Clean surface tension & $\sigma_0$ & N/m \\
    Max surfactant concentration & $\Gamma_\infty$ & mol.m$^{-2}$ &
    Dilatational viscosity (max) & $\mu_d^\infty$ & Pa.s.m \\
    Shear viscosity (max) & $\mu_s$ & Pa.s.m &
    Surface elasticity param. & $\beta_s$ & -- \\
    Surface curvature & $\kappa$ & m$^{-1}$ & & & \\
    \hline
    \multicolumn{6}{c}{\textbf{Tensors}} \\
    \hline
    Surface-excess pressure tensor & $\mathbf P$ & N/m &
    Surface identity tensor & $\mathbf I_s$ & -- \\
    Identity tensor & $\mathbf I$ & -- &
    Inviscid surface-excess tensor & $\mathbf P_0^\mathbf S$ & N/m \\
    Isotropic pressure tensor & $\mathbf P_0^\mathbf X$ & N/m$^2$ &
    Bulk viscous stress tensor & $\mathbf P_{\mu}^{\mathbf X}$ & N/m$^2$ \\
    Rate of deformation tensor & $\mathbf D$ & s$^{-1}$ &
    Surface viscous stress tensor & $\mathbf P_\mu^\mathbf S$ & N/m \\
    Surface rate of deformation tensor & $\mathbf D_s$ & s$^{-1}$ & & & \\
    \hline
    \multicolumn{6}{c}{\textbf{Vectors}} \\
    \hline
    Two-dimensional surface & $\mathbf S$ & m$^2$ &
    Three-dimensional space & $\mathbf X$ & m$^3$ \\
    Interface position & $\mathbf x_f$ & m &
    Normal vector & $\mathbf n$ & -- \\
    Surface lineal force & $\mathbf p$ & N/m$^2$ &
    Inviscid surface lineal force & $\mathbf p_0^\mathbf S$ & N/m$^2$ \\
    Surface surface viscous force & $\mathbf F_v$ & N/m$^2$ &
    Dilatational surface viscous force & $\mathbf F_v^d$ & N/m$^2$ \\
    Shear viscous force & $\mathbf F_v^s$ & N/m$^2$ & & & \\
    \hline
    \multicolumn{6}{c}{\textbf{Scalars}} \\
    \hline
    Thermodynamic pressure & $p$ & N/m$^2$ &
    Surface divergence & $\nabla_s \cdot$ & m$^{-1}$ \\
    Gradient & $\nabla$ & m$^{-1}$ &
    Divergence & $\nabla \cdot$ & m$^{-1}$ \\
    Bulk material derivative & $\dfrac{\rm D}{{\rm D}t}$ & s$^{-1}$ &
    Surface material derivative & $\dfrac{\rm D_s}{{\rm D}t}$ & s$^{-1}$ \\
    Temperature & $T$ & K &
    Heaviside function & $\mathcal H$ & -- \\
    Universal gas constant & $\mathcal R$ & -- &
    Coupling parameter & $a$ & -- \\
    Surface-viscous tension & $\sigma_{vis}$ & N/m & & & \\
    \hline
    \multicolumn{6}{c}{\textbf{Dimensionless numbers}} \\
    \hline
    Reynolds number & $Re$ & -- &
    Capillary number & $Ca$ & -- \\
    Peclet number & $Pe$ & -- &
    Boussinesq numbers & $Bq_s, Bq_d$ & -- \\
    Weber number & $We$ & -- &   Bond number & $Bo$ & -- \\ 
    \hline
    \end{tabular}
    
    \label{tab:var_classification}
\end{table*}
\section{Introduction}
A two-dimensional deforming interface separating two-phase fluid flows plays a key role in transmitting stress to the bulk phases. In response, the interface may stretch, compress or shear. Constitutive assumptions are commonly devised to describe the interfacial stress transmissivity of the two phases \cite{wasan1992interfacial, slattery2007interfacial, leal2007advanced}. The simplest interface is an extensible, inviscid surface with constant surface tension. This leads to a non-zero normal stress at the interface. The normal stress is directly related to the pressure jump across the interface and the local curvature of the surface. The first level of complexity arises when surface-active agents are present at the interface  \citep{manikantan2020surfactant}. These agents locally reduce the surface tension. The varying surface tension due to these agents generates tangential stress. This phenomenon is known as Marangoni stress. Furthermore, the presence of these agents leads to a complex structured surface that responds to additional stress upon deformation, leading to studies on interfacial rheology \citep{jaensson2018tensiometry, jaensson2021computational, matar2002surfactant, karapetsas2011surfactant}. 


The origins of interfacial rheology can be traced back to the writings of Descartes and Rumford, as outlined by Lord Rayleigh \citep{strutt1890iv}. One of the first systematic experimental studies was conducted by Plateau \citep{plateau1873statique}, who measured the damping of an oscillating magnetic needle at both clean and surfactant-covered interfaces. Marangoni \cite{marangoni1972principle} later modified Plateau's experiment by introducing a floating disc at the interface. He elucidated that the damping was due to surfactant concentration gradients that generated tangential stress, now known as Marangoni stress. Building upon this, Rayleigh replaced the disc with a ring to minimise the concentration gradients. He showed that damping was observed only because of the reduction in the contact surface area, making it one of the first recognitions of surface viscosity. Boussinesq \cite{boussinesq1913existence} later formalised these concepts by introducing surface shear and dilatational viscosities to explain phenomena such as the observation by Lebedev \cite{lebedev1916stokes} and Silvey \cite{silvey1916fall} that contaminated bubbles rise through a liquid like solid spheres. Finally, Scriven extended Boussinesq's theory by incorporating surface momentum conservation equations on a two-dimensional manifold and formulated a constitutive model for a Newtonian interface, known as the Boussinesq–Scriven constitutive relation.

The difficulties in the calculation of surface rheology in the literature inspire the development of a robust numerical method employing surfactant-dependent flows with Marangoni and surface viscous stresses. A summary of the different numerical methods used for surface viscous interfacial flows is shown in Table \ref{table-1}, which shows that no robust Direct Numerical Simulation has yet been implemented for surface-viscous stresses in its full form. Pozrikidis \cite{pozrikidis1994effects} used Boundary element method for a constant surface tension and surface viscosity. However, this method is computationally expensive; therefore, the steady-state solutions were not evaluated. Gounley et al., \cite{gounley2016influence} later extended the method to obtain the steady state deformation and successfully compared with the theory for Flumerfelt \cite{flumerfelt1980effects}. Singh and Narsimhan \cite{singh2020deformation} utilised a two-dimensional boundary element method to separately analyse the dilatational and shear surface viscosities. They also utilised their method to analyse different applications involving surface viscosity \citep{singh2021impact, singh2022numerical, singh2023impact, singh2024effect}. Reusken et al., \cite{reusken2013numerical} formalised a three-dimensional Level-set method where a drop is under a laminar flow and Stokes regime. Although they formalised the method for  both shear and dilatational surface viscosities, their validation is strongly based on the evaluation of dilatational surface viscosity. The above works are discussed in the Stokes regime. For the unsteady Navier-Stokes regime, \cite{ubal2005influence} developed an arbitrary Lagrangian-Eulerian (ALE) method to study the effects of surface viscosity on two-dimensional Faraday waves. Moreover, \cite{wee2020effects} developed a sharp interface method based on ALE to study the role of interfacial rheology in the pinch-off of liquid threads. \cite{luo2019influence} developed a three-dimensional Front-tracking method that includes both surfactant and surface rheological effects on a drop under shear flows. These methods were developed for specific applications, such as low Reynolds numbers, axial symmetry, or low deformation of drops without breakup. In this study, we aimed to develop a three-dimensional generalised Navier-Stokes solver with a complex interface consisting of surfactant-covered flows with interfacial rheology. 

Numerical studies of complex interfacial flows have focused on the accurate tracking or capturing of the moving interface when the surface tension is constant. Some popular front-capturing methods are volume of fluid \cite{gopala2008volume}, phase field \cite{lamorgese2011phase}, Lattice Boltzmann \cite{aidun2010lattice}, and level set methods \cite{sethian2003level}. In contrast, the front-tracking method \cite{unverdi1992front} employs a separately tracked Lagrangian grid of interface elements, which provides an accurate representation of the interface position and a robust and accurate calculation of the surface tension forces. Hybrid methods have also been developed, wherein the advantages of one method are retained while avoiding the inconvenient aspects of the other. One such method is the Level Contour Reconstruction Method (LCRM) \cite{shin2002level}, which retains the front-tracking interface for the accurate calculation of the interface and also retains the ease with which topological coalescence and rupture of the interface are handled by the level-set method. 

In a full three-dimensional (3D) simulation, the most challenging task is to solve the two-dimensional (2D) surface viscous stress confined on the interface and transmit these effects to the 3D bulk phase. Additionally, a correct representation of the surface viscous stress is crucial for the implementation of a generalised interfacial flow, irrespective of its spherical or axial symmetries. Surface viscous stresses act as 2D momentum diffusion on the interface which is subjected to extreme topological changes, such as breakup, coalescence, or impulse. The momentum diffusion is handled by the Lagrangian elements, where a secondary two-dimensional viscous stress is explicitly solved for the surface shear viscosity, and by the one-fluid formulation, the surface viscous stress effects are transmitted to the bulk sub-phase. To account for surface momentum diffusion, information on the neighbouring Lagrangian elements is required. This implies that algorithms for bookkeeping the connected elements must be developed. Classical front-tracking methods have developed such algorithms which are primarily difficult to scale and parallelise. In our hybrid LCRM, bookkeeping is avoided, and the interface is reconstructed using the level-set function. This results in a simpler code structure compared to the classical front-tracking approach and parallelises efficiently. The work presented here is developed in the in-house code BLUE, a massively parallel multiphase flow solver, using newly developed numerical techniques for interfacial stresses that are suitable for distributed processing.        

\begin{table}[h!]
\centering
\footnotesize
\caption{Summary of numerical methods for surface viscous interfacial flows. All these methods are implemented to include dilatational and shear surface viscosities using different interface-tracking methods and mesh structures. In this study, a hybrid level-set/front-tracking method is proposed to include surface rheology in unsteady Navier–Stokes solutions. 2D and 3D represent two- and three-dimensional computational domains, respectively. \\}

\begin{tabular}{>{\raggedright\arraybackslash}p{4cm} 
                >{\raggedright\arraybackslash}p{2cm} 
                >{\raggedright\arraybackslash}p{4cm} 
                >{\raggedright\arraybackslash}p{2cm}}
\toprule
\textbf{Interface tracking method} & \textbf{Dimensions} & \textbf{General remarks} & \textbf{Reference} \\
\midrule\midrule
Boundary element method & 2D (cartesian and cylindrical) & This method is applicable for low Reynolds number and the simulations are carried out below or equal to the critical capillary number.  & \cite{singh2021impact, singh2022numerical} \\ \\
                        & 3D                           & This is also applicable to low Reynolds number. & \cite{pozrikidis1994effects, gounley2016influence}\\
\midrule
Level-set method        & 3D                           & For low Reynolds number and validated only for dilatational surface viscosity & \cite{reusken2013numerical} \\
\midrule
Arbitrary Lagrangian Eulerian & 2D (cylindrical, cartesian) & Unsteady Navier-Stokes regime but the shear and dilatational surface viscosities are indistinguishable &\cite{ubal2005influence, wee2020effects, wee2023effects}\\
\midrule
Front-tracking & 3D & Unsteady Navier-Stokes regime but the implementation can not be extended for the fragmentation/coalescence of free surface & \cite{luo2019influence}\\
\midrule
Level-set based front-tracking & 3D & Unsteady Navier Stokes regime with the capabilities to handle topological changes of the interface due to breakup and coalescence & Present method \\
\bottomrule
\end{tabular}
\label{table-1}
\end{table}

\section{Problem formulation and numerical implementation}
In this section, we discuss the numerical implementation of surface viscosity, dilatational and shear, in the context of the Level Contour Reconstruction Method (LCRM), our hybrid level-set front tracking method, in Section \ref{sec_LCRM}. Next, we describe the interfacial mechanics associated with a Boussinesq-Scriven constitutive model 
in section \ref{BS}. Subsequently, the governing equations for two-phase flows with surfactant-driven elastic and viscous interfaces are discussed in Section \ref{governing_equations}. Finally, we discuss the numerical implementation of surface viscous forces in detail in Section \ref{sv_ce}. The discretisation and implementation of surfactant conservation equations are briefly discussed in Section \ref{surf_ce}. More details on the implementation of the surfactant conservation equations can be found elsewhere \cite{Shin2018ASurfactant}. 

\subsection{Level Contour Reconstruction Method}\label{sec_LCRM}
The motion of the interface is tracked using the LCRM, which is a hybrid approach for two-phase flow simulations that combines the geometric accuracy of the classical front-tracking method with the automatic topology handling capabilities of the level-set front-capturing method. 
\begin{figure}[h!]
    \centering
    \includegraphics[width=0.8\linewidth]{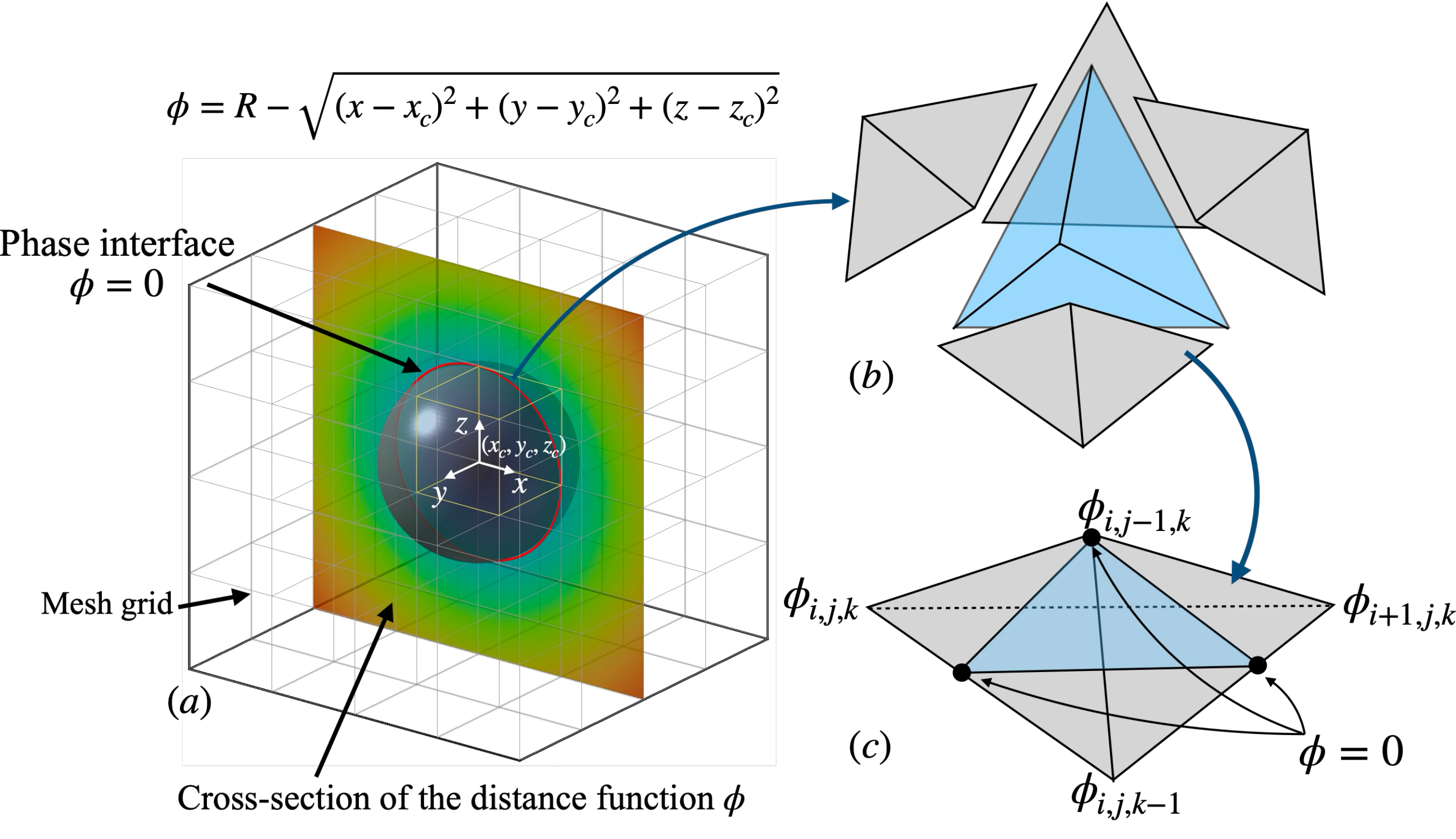}
    \caption{General description of the LCRM}
    \label{LCRM}
\end{figure}
%
%
A schematic representation of the LCRM in 3D is shown in figure \ref{LCRM}. Starting from a set of Lagrangian interface elements, a distance function field is reconstructed on an Eulerian grid \cite{shin2009hybrid}. Implicit connectivity among Lagrangian elements is achieved by subdividing each Eulerian cell into tetrahedra (see figure. \ref{LCRM}). This enables a cell-wise reconstruction of the zero-isocontour surface and the identification of its intersection with the Lagrangian surface. This tetra-marching procedure guarantees that only one unique isocontour surface is resolved for each reconstruction cell. The reconstruction is typically performed at every $25$ time steps, with the frequency chosen such that the maximum element displacement per step does not exceed the smallest Eulerian cell width. Regular reconstruction prevents element distortion, avoids dispersion or clustering owing to interface deformation, and naturally accommodates topological changes as new elements inherit the topology of the reconstructed distance function. Higher-order reconstructions further improved the interface accuracy. The method maintains a mass conservation error below $0.1\%$ and significantly improves surface tension evaluation, thereby suppressing spurious parasitic currents. 

The LCRM is integrated into our code BLUE, which is massively parallelised (tested on up to 131072 processors). The code is suitable for the direct simulation of incompressible flows with surface tension-driven interfaces. A parallel hybrid multigrid/GMRES algorithm efficiently solves the pressure Poisson equation even at very high density ratios of $O(10^4)$. The parallelisation of the LCRM with a message-passing interface and domain decomposition is straightforward because all interface operations are local to an element and its local region of grid cells, and the characteristic features of the LCRM are inherited in each subdomain. Various modules of BLUE are dedicated to a wide variety of multiphase scenarios, and the code has been rigorously tested on a suite of multiphase benchmark problems, as well as academic research problems in the chemical and pharmaceutical industries. 
The BLUE code has been tested for surfactant-laden flows in the bulk and at the interface, moving contact lines, deformable solids, and stratified fluids \cite{panda2024marangoni,shin2020interaction, shin2018direct, abdal2024pairwise}. In this study, the main goal is to present DNS for surface-viscosity-driven interfacial flows.

\subsection{Boussinesq-Scriven Newtonian surface viscous fluid} \label{BS}
\begin{figure}[h!] 
    \centering
    \includegraphics[width=0.85\linewidth]{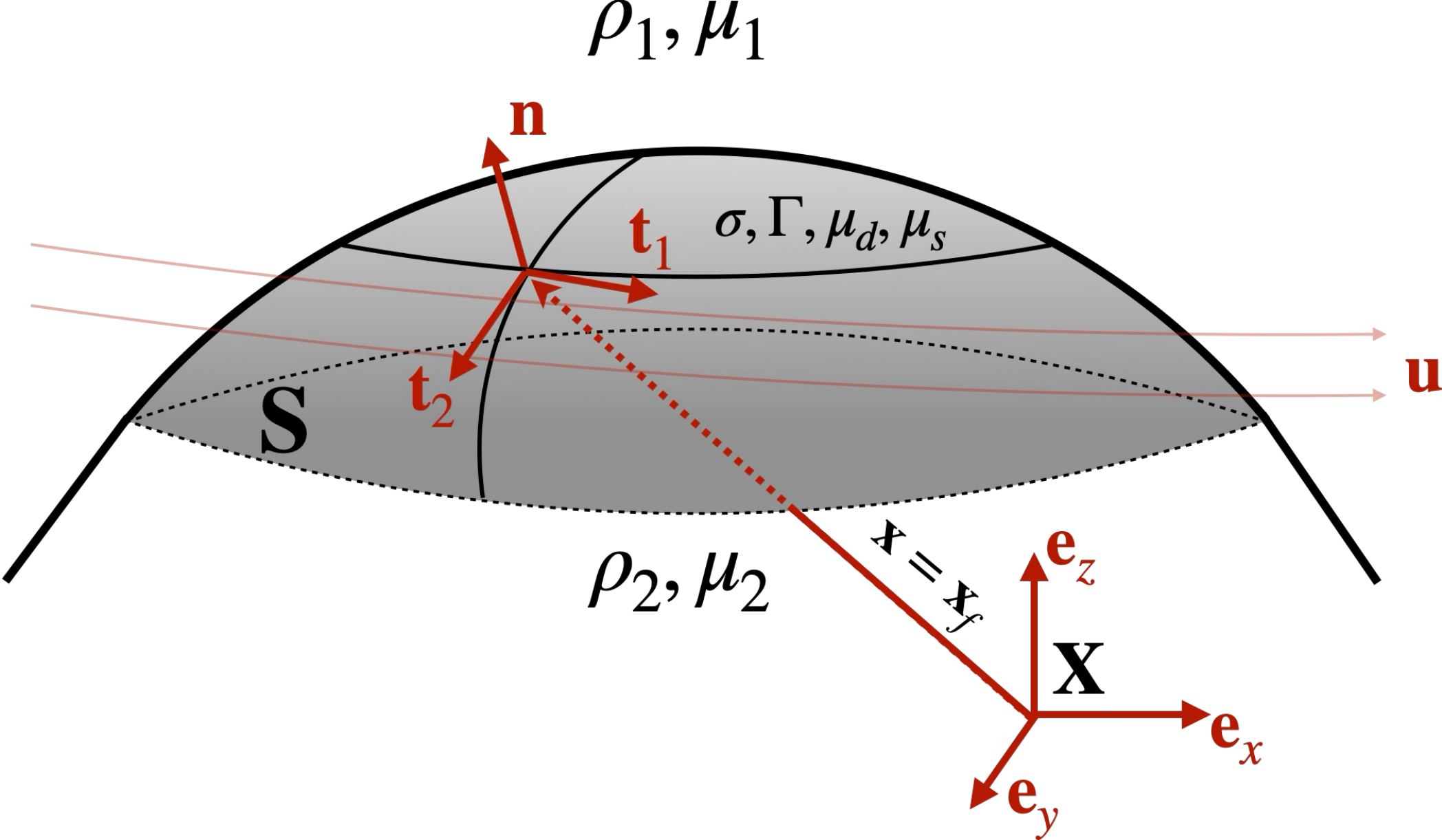}
\caption{Graphical representation of an extrinsic viewpoint of the Boussinesq-Scriven surface geometry: the three-dimensional space, $X$ is of a Cartesian basis, referring to the two-dimensional manifold, $S$, at $\mathbf x = \mathbf x_f$. A local coordinate system of basis $\mathbf n$, $\mathbf t_1$, and $\mathbf t_2$ defines the manifold, which we call the \textit{interface}. From an intrinsic viewpoint, the observer is unaware of the normal, $n$, and describes the manifold as a two-dimensional space. The intrinsic and extrinsic viewpoints converge if and only if $\text{span}(\mathbf t_1, \mathbf t_2)$ is the basis of the two-dimensional intrinsic system.  The interface differentiates the two phases as subscripts $1$ and $2$, and has densities and viscosities of $\rho_{1}, \rho_2$ and $\mu_1, \mu_2$, respectively. The surface properties are surface tension $\sigma$, surface shear viscosity $\mu^S_s$, and surface dilatational viscosity $\mu^S_d$. Here, $\mathbf u$, is the velocity defined in $\mathbf X-$ space.}
\label{fig:method_0}
\end{figure}

Surface geometry can be described from an intrinsic or extrinsic perspective. In the intrinsic formulation, the surface is treated as a 2D manifold ($\mathbf S$), with the geometry defined from the viewpoint of an observer constrained to the surface. In contrast, the extrinsic viewpoint considers the manifold $\mathbf S$ to be embedded in a 3D space ($\mathbf X$). The Boussinesq-Scriven constitutive model has been derived using both perspectives in the literature (see Table \ref{tab_2}).
The manifold, $\mathbf S$, represents the interface that separates two fluid phases with densities $\rho_1, \rho_2$, and viscosities, $\mu_1, \mu_2$, in 3D space, $\mathbf X$. For instance, within the framework of the intrinsic viewpoint, the interface can be defined based on the basis $(\mathbf t_1, \mathbf t_2)$, as shown in figure \ref{fig:method_0}. Defining it in such a framework reduces the problem of solving the surface kinematics without knowing the background fluid flow. From an extrinsic viewpoint, the surface is defined within a 3D framework, $\mathbf X$, and the interfacial position is determined by $\mathbf x = \mathbf x_f$ and the orientation by $\mathbf n$. Because we employ a hybrid level-set-based Lagrangian interface tracking, we derive the Boussinesq-Scriven formulation from an extrinsic viewpoint. 

The deforming interface is subjected to two types of forces: (1) surface body forces (areal), such as gravitational or electromagnetic forces, which originate in $\mathbf X$ and are transmitted to $\mathbf S$ by the two phases; and (2) surface contact forces, such as capillary, Marangoni, and surface viscous forces, which originate on $\mathbf S$ due to the contiguity of the interfacial elements. From an intrinsic viewpoint, the observer is unaware of the areal force. On the other hand, in the extrinsic view, the areal forces are generally added as volumetric forces in the momentum equation. However, the main contribution to the interfacial forces is associated with the surface linear forces; this is the main subject of the present study. 

We derive the surface lineal force ($\mathbf p$) by operating the surface divergence ($\nabla_s \cdot (\cdot) $) of the surface-excess pressure tensor, $\mathbf P$, such that, 
\begin{equation}
    \mathbf p = \nabla_s \cdot \mathbf P.
\end{equation}
Here, $\nabla_s \cdot (\cdot) = (\mathbf I_s)\nabla \cdot(\cdot)$ is computed by projecting the divergence operator on $\mathbf S$ using the surface projection tensor $\mathbf I_s = \mathbf I - \mathbf n \mathbf n$, where $\mathbf n$ is the normal vector to the interface. For an inviscid surface, the surface-excess pressure tensor, $\mathbf P_0 $, is given by, 
\begin{equation}
    \mathbf P_0^\mathbf S = \sigma \mathbf I_s,
\end{equation}
where $\sigma$ is the surface tension. 
This is similar to the pressure tensor in bulk fluid ($\mathbf P^\mathbf X$); we evaluate the isotropic pressure tensor by $\mathbf P^\mathbf X _0= -p \mathbf I$, where $p$ is the thermodynamic pressure. Analogous to the divergence of bulk pressure tensor ($\nabla \cdot (\mathbf P^\mathbf X_0)=\nabla \cdot (-p \mathbf I) = -\nabla p$) to evaluate the pressure gradient volumetric force, the ideal, inviscid, lineal force on the surface $\mathbf p_0^\mathbf S$  is evaluated as
\begin{equation}
   \mathbf p_0^\mathbf S = \nabla_s \cdot (\sigma \mathbf I_s) =(\nabla_s \cdot \mathbf I) \sigma + (-\nabla_s \cdot \mathbf{n})(\mathbf n \sigma) = \nabla_s \sigma + \kappa (\mathbf n\sigma),
\end{equation}
where $\kappa = -\nabla_s \cdot \mathbf n$, is the local surface curvature. 
The inviscid lineal force $\mathbf p^0$ has two parts: (a) the normal component in the direction $\mathbf n$, that quantifies the deformation of the interface as proportional to its local curvature $\kappa$, and (b) the surface tangential forces, driven by the heterogeneity in the surface tension ($\nabla_s \sigma$). Numerical methods for solving inviscid lineal forces are well-known in the community \cite{tryggvason2001front, shin2018hybrid, muradoglu2008front, farsoiya2024coupled}.

We draw an analogy between the bulk viscous stress tensor and the surface-excess viscous stress tensor to derive the generalised Boussinesq-Scriven constitutive equation. 
The bulk viscous stress tensor, $\mathbf P^\mathbf X_\mu$ is given by,
 \begin{equation}
     \mathbf P^\mathbf X_\mu = \left(\mu_d^\mathbf X-\dfrac{2}{3}\mu_s^\mathbf X\right)\left(\mathbf I : \mathbf D\right)\mathbf I + 2\mu_s^\mathbf X \mathbf D. 
 \end{equation}
Here, $\mu^\mathbf X_d$ and $\mu_s^\mathbf X$ are the dilatational and shear bulk viscosities, respectively, and $\mathbf D = \dfrac{1}{2}\left(\nabla \mathbf u + \nabla \mathbf u^{\rm T}\right)$ is the rate of deformation tensor. The operation $\mathbf I : \mathbf D$ reduces to $\nabla \cdot \mathbf u$, which vanishes in the incompressible limit. 
Similarly, the surface-excess viscous stress tensor, $\mathbf P^s_\mu$, is given by, 
\begin{equation}
    \mathbf P^s_\mu = \left(\mu_d^\mathbf S - \mu_s^\mathbf S\right)\left(\nabla_s \cdot \mathbf u_f\right) + 2\mu_s^\mathbf S \mathbf D_s.
\end{equation}
Here, $\mathbf u_f$ is the velocity at the interface and $\mathbf x = \mathbf x_f$. $\mu_d^\mathbf S$ and $\mu_s^\mathbf S$ are the dilatational and shear surface viscosity, and $\mathbf D_s$ is the surface deformation rate tensor. The interpretation of $\mathbf D_s$ varies in the literature owing to the differences in (a)  the nature of the problem, for example, low Reynolds flow and thin films; (b) the utility of theoretical approaches, for example, linear stability, boundary integral, and lubrication methods; and (c) the choice of coordinate system, for example, intrinsic and extrinsic viewpoints. 

\begin{table}[h!]
    \centering
    \footnotesize
    \renewcommand{\arraystretch}{1.3} 
    \setlength{\extrarowheight}{2pt}  

    \caption{Literature survey of the different formulation of surface deformation rate, $\mathbf{D}_s$}
    \label{tab_2}

    \begin{tabular}{p{0.2\textwidth}p{0.2\textwidth}p{0.5\textwidth}} 
        \hline
        \textbf{Literature} & $2\mathbf{D}_s$ & \textbf{Remarks} \\[3pt]
        \hline
        \citealp{ponce2017influence} & 
        $\nabla_S \mathbf u_S + (\nabla_S \mathbf u_S)^T$ & 
        {\bf{Intrinsic viewpoint:}} $\mathbf u^S$ is a two-dimensional tangential velocity, and $\nabla^S$ is the intrinsic surface gradient. A specific coordinate transformation from an extrinsic Cartesian coordinate to the local coordinate of the interface leads to the evaluation of $\nabla^S$ and $\mathbf{u}^S$ \citep{herrada2022stability}.  \\
         \citealp{secomb1982surface} & $\mathbf I_s \cdot \left(\nabla \mathbf u^* + (\nabla \mathbf u^*)^T \right) \cdot \mathbf I_s$ & {\bf{Extrinsic and Eulerian viewpoint:}} The authors derived a simple expression that reduces the surface deformation rate tensor as the surface projection of the bulk deformation rate tensor. Furthermore, they also prescribed that the velocity may not only be on the interface, but also at the nearest location to the interface, for example, the nearest neighbour of the interface in a Eulerian grid ($\mathbf u_f \approx \mathbf u^*$).\\
        \\ 
        \citealp{lopez1998direct} & $\nabla_s \mathbf u_s \cdot \mathbf I_s + \mathbf I_s \cdot (\nabla_s \mathbf u_s)^T $ & {\bf{Extrinsic and Lagrangian viewpoint:}} Here the authors perform surface projection of the surface gradient of surface velocity ($\nabla_s \mathbf u_s$, where $\mathbf u_s = \mathbf u \cdot \mathbf I_s$) and the symmetric part of the tensor is evaluated as the surface deformation rate tensor. In their work, they simplified the surface velocity parallel to an $xy$ plane as follows: Considering the surface velocity results in an explicit term which is non-negligible for a finite deformable surface. Readers are suggested to refer  \ref{appA} for more details. \\
        \citealp{jaensson2021computational, scriven1960dynamics} & $\nabla_s \mathbf u_f \cdot \mathbf I_s + \mathbf I_s \cdot (\nabla_s \mathbf u_f)^T$ & {\bf{Extrinsic and Lagrangian viewpoint:}} The authors have projected the surface gradient of the interfacial velocity and its transpose to evaluate the $\mathbf D_s$. Thus, the surface gradient of the interfacial velocity is calculated on the Lagrangian surface, $\mathbf x = \mathbf x_f$, and the projection is operated to reduce the tensor on the surface from an extrinsic viewpoint. \\
        \hline
    \end{tabular}

\end{table}

A detailed summary of the various viewpoints and their formulations is presented in Table \ref{tab_2}. From an intrinsic viewpoint \cite{ponce2017influence, herrada2022stability}, $\mathbf D_s$ is similar to the 2D representation of $2\mathbf D = \nabla \mathbf u + (\nabla \mathbf u)^T$. The rate of deformation tensor is evaluated, where a sophisticated coordinate transformation is implemented to utilise the background flow field. Such methods are easily implemented for 2D (or axisymmetric cylindrical) problems, where the surface is reduced to a one-dimensional (1D) vector. From the extrinsic viewpoint, the background velocity is interpolated at the interface as $\mathbf u(\mathbf x_f) = \mathbf u_f$. In a 2D problem, the surface gradient is easily obtained because the normal and tangent vectors are calculated as a function of the interface height function. Secomb and Skalak \cite{secomb1982surface} derived a simpler method in which the velocity interpolation at the interfacial position is not required. Instead, they suggested directly operating a surface projection on the nearest Eulerian grid cell, where $\nabla \mathbf u$ is already calculated in the Eulerian field. We refer to this method as the extrinsic-Eulerian viewpoint. Scriven \cite{scriven1960dynamics} derived a surface viscous momentum equation in the reference frame of $\mathbf S$, which is integrated with background fluid flow information; this method is called the extrinsic Lagrangian viewpoint. Because our numerical method is based on a hybrid formulation of the level set, where the extrinsic viewpoint is accessible, and front tracking, where the Lagrangian viewpoint is implementable, the extrinsic-Lagrangian viewpoint is a viable choice for formulating the shear surface viscous forces. Other formulations have also been implemented in the literature and are summarised in Tables \ref{tab_2} and \ref{appA}.   

In the next subsection, the subscripts and superscripts on the interfacial velocity, as provided in Table \ref{tab_2}, are dropped. In the one-fluid formulation, the interfacial velocity is obtained by interpolating the velocity from the Eulerian grid to the Lagrangian grid such that $\mathbf u_f = \mathbf u(\mathbf x=\mathbf x_f)$. 

\subsection{Governing equations} \label{governing_equations}
The governing equation for the mass and momentum conservation of a two-phase, incompressible, and surface-viscous fluid, is described in a single-fluid formulation given by, 
\begin{equation}
    \nabla \cdot \mathbf u = 0,
\end{equation}
\begin{align}
    \rho \dfrac{{\rm D}\mathbf u}{{\rm D}t} &= \nabla \cdot (-p\mathbf I + 2\mu \mathbf D) + \rho \mathbf f \nonumber \\
    &\quad + \int_{A'} \nabla_s \cdot \left[\left(\sigma + (\mu_d - \mu_s)(\nabla_s \cdot \mathbf u)\right) \mathbf I_s\right] \delta(\mathbf x - \mathbf x_f) ~ dA' 
    + \int_{A'} \nabla_s \cdot (2\mu_s \mathbf D_s) ~ \delta(\mathbf x - \mathbf x_f) ~ dA'.
    \label{mom}
\end{align}
Here, $\rm D$ is the material derivative (${\rm D(\cdot)/{\rm D} }t \equiv \partial(\cdot)/\partial t + \mathbf u \cdot \nabla(\cdot)$), $\mathbf f$ is the volumetric force vector, $\nabla_s\cdot (\nabla_s)$ is the surface divergence (gradient), $\mathbf I ~(\mathbf I_s)$ is the identity (surface identity) tensor, $\delta$ is the Kronecker delta function, which is non-zero only at the interface $\mathbf x = \mathbf x_f$. The surface material properties are the surface tension, $\sigma~ (\rm{kg. s^{-2}})$, dilatational surface viscosity, $\mu_d~ (\rm{kg. s^{-1}})$, and shear surface viscosity, $\mu_s~ (\rm{kg.s^{-1}})$. The bulk material properties are the single-fluid formulated density, $\rho ~(\rm{kg.m^{-3}})$ and viscosity, $\mu~(\rm {kg .m^{-1}.s^{-1}})$. These material properties are given by, 
\begin{equation}
\begin{aligned}
    \rho(\mathbf x, t) &= \rho_1 \mathcal H(\mathbf x, t) + \rho_2 \left(1 - \mathcal H(\mathbf x, t)\right), \\
    \mu(\mathbf x, t)  &= \mu_1 \mathcal H(\mathbf x, t) + \mu_2 \left(1 - \mathcal H(\mathbf x, t)\right).
\end{aligned}
\end{equation}
$H(\mathbf x, t)$ is the Heaviside function, which is $1$ for phase $1$ and $0$ otherwise. 

We considered that the interface is contaminated with surfactants. This implies that the surface tension changes locally and is governed by the packing of surfactants at the interface. The surfactant concentration at the interface is $\Gamma, ~(\rm mol.m^{-2})$. For simplicity, we considered the surfactants to be insoluble. This indicates that the ad/desorption of surfactants from the bulk is negligible. The surface transport equation governing the surfactant transport is given by, 
\begin{equation}
    \dfrac{\rm{D_s} \Gamma}{{\rm{D}} t} = D_\Gamma \nabla_s^2\Gamma,
\end{equation}
where ${\rm{D}_s}(\cdot)/{\rm D}t = \partial(\cdot)/\partial t + \nabla_s ( (\cdot) \mathbf u_s)$ is the surface material derivative and $D_\Gamma$ is the diffusivity of the surfactants. Under the assumption of an insolubility limit, the Langmuir-Szyskowski nonlinear equation of state can be used to couple surfactant dynamics to hydrodynamics:  
\begin{equation}
    \sigma(\Gamma) = \sigma_0 + \text{R}T\Gamma_\infty ~ \ln \left(1- \dfrac{\Gamma}{\Gamma_\infty}\right)
.\end{equation}
Here, $\sigma_0$ is the surface tension of the clean interface ($\Gamma = 0$), $R$ is the universal gas constant, $T$ is the temperature, and $\Gamma_\infty$ refers to the maximum packing of the surfactant on the interface. Surface rheological effects arise when surfactants deform against themselves at the interface. Intuitively, the surface viscosities are a function of $\Gamma$. Although it is still an open question to predict $\mu_s(\Gamma)$ and $\mu_d(\Gamma)$, we choose a simple model in this work:
\begin{equation}
    \mu_s(\Gamma) = \mu_s^\infty \left(\frac{\Gamma}{\Gamma_\infty}\right)^a, ~~ \mu_d(\Gamma) = \mu_d^\infty \left(\frac{\Gamma}{\Gamma_\infty}\right)^a,
\end{equation}
where $\mu_s^\infty$ and $\mu_d^\infty$ are the surface shear and dilatational viscosities, respectively, as $\Gamma \rightarrow \Gamma_\infty$; $a$ is a numerical parameter, which is either $0$ (constant surface rheology) or $1$ (linearly varying with $\Gamma$). More sophisticated models can be implemented in future work.  

A coupling equation that inherently satisfies the surface-viscous-tension stress condition is then required to close the above set of partial differential equations. The conservation of mass implies that the material derivative of density, ${\rm D}\rho / {\rm D}t = 0$. For discontinuous phases, this is simplified to the material derivative of the Heaviside function, ${\rm D}H/Dt = 0$. Hence, the interface satisfies the stress condition, is implicitly tracked by the Heaviside function, and is advected by the material motion of the background fluid. 

The interfacial elements are advected in Lagrangian fashion by integrating, 
\begin{equation}
    \frac{d\mathbf x_f}{dt} = \mathbf u_f,
\end{equation}
with a second-order Runge-Kutta method, where the interface velocity $\mathbf u_f$ is interpolated from the Eulerian velocity. A well-known projection method on a staggered MAC mesh is used to solve for the fluid velocity and pressure. A second-order ENO scheme is used for the convective terms. A more detailed description of the procedure for solving the momentum equation can be found in \cite{shin2002modeling, shin2009hybrid}.

\begin{figure}[h!]
    \centering
    \includegraphics[width=0.7\linewidth]{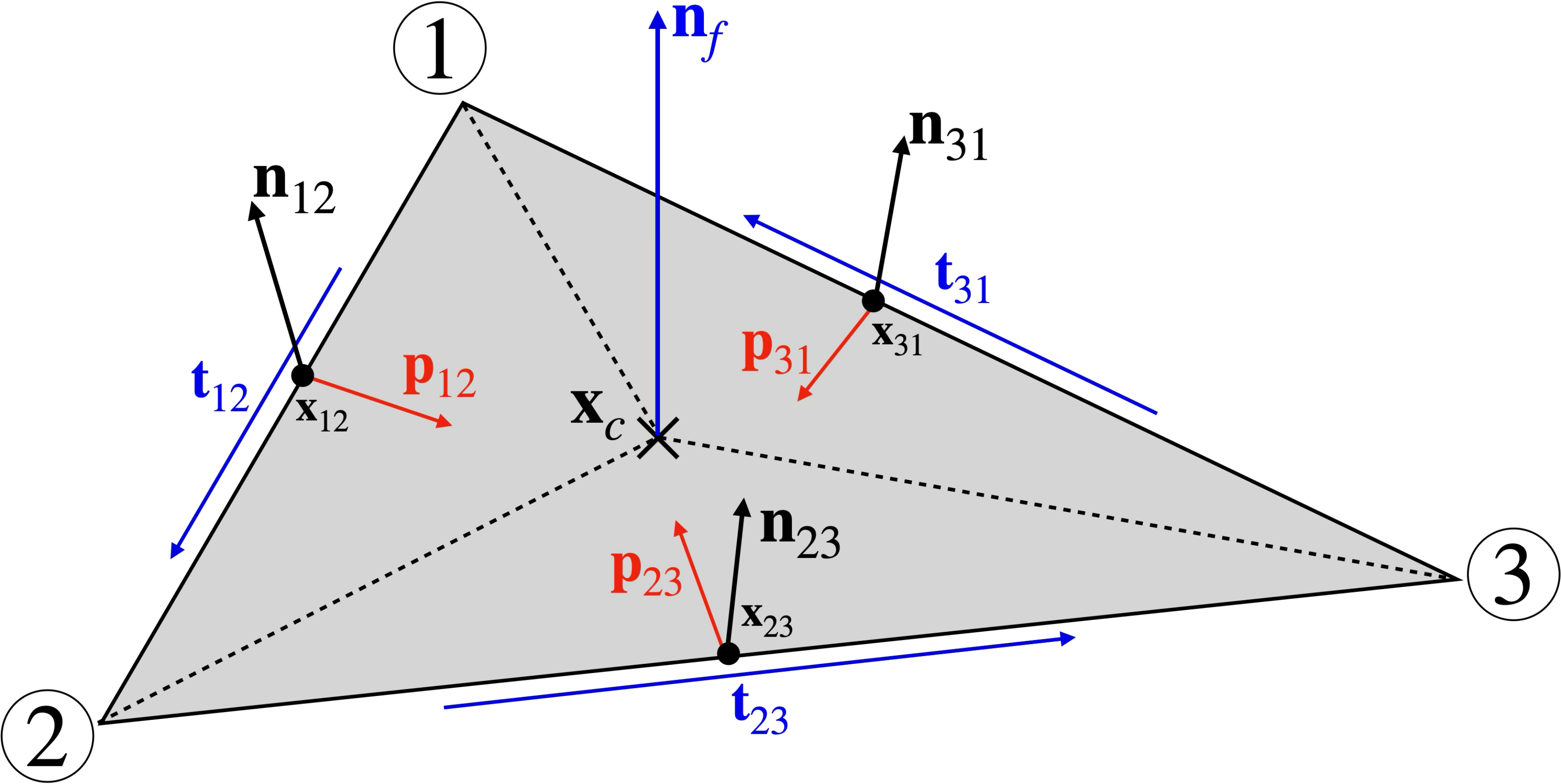}
    \caption{\textbf{Description of geometrical information for an individual interface element: normal, binormal, tangent vectors at the edges of the element as well as the interface normal at the centre of the element.}}
    \label{Surface information}
\end{figure}

\begin{figure}[h!]
    \centering
    \includegraphics[width=1.0\linewidth]{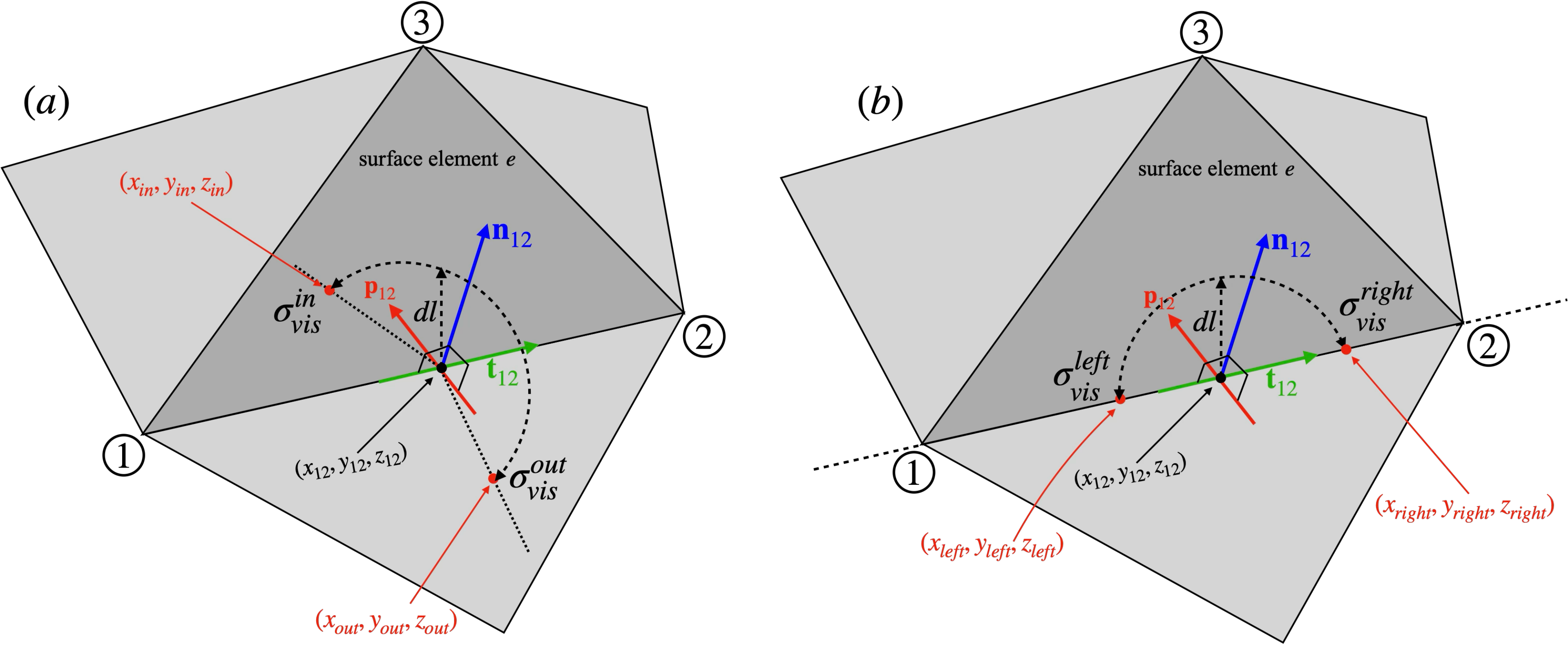}
    \caption{\textbf{Description of geometrical information for an individual interface element: normal, binormal, tangent vectors at the edges of the element as well as the interface normal at the centre of the element.}}
    \label{probing_technique}
\end{figure}

\subsection{Surface viscous conservation equations} \label{sv_ce}
In the momentum equation \eqref{mom}, the surface viscous force, $\mathbf F_v$, is given by, 
\begin{equation}
\mathbf F_v =
\underbrace{\int_{A'} \nabla_s \cdot \left[(\mu_d - \mu_s)(\nabla_s \cdot \mathbf u)\mathbf I_s\right] 
\delta(\mathbf x - \mathbf x_f) \, dA'}_{\mathbf F_v^d}
+ \underbrace{\int_{A'} \nabla_s \cdot (2\mu_s \mathbf D_s) \,
\delta(\mathbf x - \mathbf x_f) \, dA'}_{\mathbf F_v^s}.
\label{sv1}
\end{equation}
The surface viscous force can be decomposed into two components: (1) dilatational surface viscous force ($\mathbf F_v^d$) and (2) shear surface viscous force ($\mathbf F_v^s$). We simplify $\mathbf F_v^d$ by introducing a surface-viscous tension, $\sigma_{vis}$, given by,
\begin{equation}
    \sigma_{vis} = (\mu_d - \mu_s)(\nabla_s \cdot \mathbf u).
    \label{sv2}
\end{equation}
The dilatational surface viscous force on each element $e$ can then be simplified as, 
\begin{equation}
\mathbf F_v^d = \int_{A_e} \nabla_s \cdot (\sigma_{vis} \mathbf I_s) ~\delta(\mathbf x-\mathbf x_f)~ 
dA_e =
\underbrace{\int_{A_e} \sigma_{vis} \kappa \mathbf n ~\delta (\mathbf x - \mathbf x_f)~dA_e}_{\mathbf{F}_n}
+ \underbrace{\int_{A_e} \nabla_s \sigma_{vis} ~ \delta (\mathbf x - \mathbf x_f)~dA_e}_{\mathbf{F}_s},
\label{sv3}
\end{equation}
where $A_e$ is the area of element $e$. One such surface element is shown in figure \ref{Surface information}. The element is a triangle, where the vertices are named as $\textcircled{1}, \textcircled{2}$, and $\textcircled{3}$. The midpoints of the edges are $\mathbf x_{12}, \mathbf x_{23}, $ and $\mathbf x_{31}$ where the normals are defined by $\mathbf n_{12}, \mathbf n_{23}, $ and $\mathbf n_{31}$, respectively. The binormal and tangential vectors are designated as $\mathbf p$ and $\mathbf t$. The centroid of the element is $\mathbf x_c$ and the normal of the element is $\mathbf n_f$.  
Equation \eqref{sv3} reduces to a varying surface-viscous tension on the interface, resulting in a normal component ($\mathbf F_n$) and tangential component ($\mathbf F_t$). This surface viscous tension force is a source term in the momentum equation which can be written analogously to the surface tension forces as, 
\begin{equation}
    \mathbf F_v^d = \int _c \sigma_{vis} \mathbf p' dl = \mathbf F_n + \mathbf F_s.
\end{equation}
Therefore, the formulation for evaluating $\mathbf F_v^d$ is similar to the calculation of surface tension forces in our hybrid level set/front-tracking method. For the sake of completeness, we discuss the implementation of $\mathbf F_v^d$ in this section.

In our hybrid formulation, $\mathbf F_n$ can be obtained using the discrete curvature $\kappa_H$:
\begin{equation}
    \mathbf F_n = \int_{A_e} \sigma_{vis} \kappa \mathbf n ~\delta(\mathbf x- \mathbf x_f) ~dA_e = \sigma_{vis}\kappa _H \nabla \mathcal I. 
\end{equation}
Since, $\sigma_{vis}$ is a varying coefficient along the interface, we computed $\sigma_{vis}\kappa_H$ as a single field distribution, given by, 
\begin{equation}
    \sigma_{vis} \kappa_H = \frac{\mathbf F_L \cdot \mathbf G}{\mathbf G \cdot \mathbf G},
\end{equation}
where, 
\begin{equation}
    \mathbf F_L = \int_{A_e} \sigma_{vis} \kappa_f \mathbf n_f ~\delta(\mathbf x- \mathbf x_f) ~dA_e,
\end{equation}
\begin{equation}
    \mathbf G = \int_{A_e}  \mathbf n_f ~ \delta (\mathbf x- \mathbf x_f)dA_e.
\end{equation}
$\kappa_f$ is twice the mean interface curvature obtained from the Lagrangian interface structure. The geometric information, unit normal, $\mathbf n_f$, and area of the element, dA, are computed directly from the Lagrangian interface and then distributed onto an Eulerian grid using a discrete Dirac distribution. The discrete field of these quantities (say $X_f$) can be computed by distributing $X_f$ to the Eulerian grid as
\begin{equation}
    M_{sijk} = \sum_f X_f D_{ijk}(\mathbf x_g) \Delta A_f,
    \label{peskin1}
\end{equation}
where $\Delta A_f$ is the element area, and $D_{ijk}$ is the discrete direc distribution. For a given interface element at a position $\mathbf x_f = (x_f, y_f, z_f)$, we use tensor product suggested by Peskin and McQueen \cite{peskin1995general}, 
\begin{equation}
    D_{ijk}(\mathbf x_f) = \frac{\delta (x_f/h_x -i)~\delta (y_f - j) ~\delta(z_f - k)}{h_x h_y h_z},
    \label{peskin2}
\end{equation}
where $h_x, h_y, $ and $h_z$ are the dimensions of an Eulerian grid cell and
\begin{equation}
\delta(r) =
\begin{cases}
\delta_1(r), & |r| \leq 1, \\
\dfrac{1}{2} - \delta_1(2 - |r|), & 1 < |r| < 2, \\
0, & |r| \geq 2
\end{cases}
\label{peskin3}
\end{equation}

and 
\begin{equation}
    \delta_1(r) = \frac{3-2|r| + \sqrt{1 + 4|r| -4r^2}}{8}.
    \label{peskin4}
\end{equation}
Using equations \eqref{peskin1}, \eqref{peskin2}, \eqref{peskin3}, and \eqref{peskin4}, the geometric information computed on the Lagrangian interface is distributed over a narrow width of $3-4$ grid cells around the interface. 

To evaluate the tangential component, $\mathbf F_t$, the surface gradient of the surface-viscous tension must be evaluated. The surface-viscous tension gradient at the interface is further decomposed into its $\mathbf p$ and $\mathbf t$ components:
\begin{equation}
    \nabla_s \sigma_{vis} = (\nabla_s \sigma_{vis})_p \mathbf p + (\nabla_s \sigma_{vis})_t \mathbf t 
\end{equation}
We used a probing technique originally introduced by Udaykumar et al. \cite{udaykumar1999computation}. The schematics for a general implementation procedure of the probing technique to compute the surface gradient of $\sigma_{vis}$ in both the $\mathbf p$ and $\mathbf t$ directions are shown in figure \ref{probing_technique}. An example is described to construct the surface gradient in the $\mathbf p$ direction at the centre between nodes \textcircled{1} and \textcircled{2}: First, a probe point is constructed $(x_{12}, y_{12}, z_{12})$ and define a probe distance $dl$ (usually equivalent to the grid size), in the normal direction $\mathbf n_{12}$. Because the interface is represented by the zero isocontour of the distance function, $\phi$, we can locate the probe point on the interface where $\phi = 0$. We then interpolate the surface-viscous tension, $\sigma_{vis}$, at the two points on either side of the interface from the probe point, that is, $\mathbf x_{\rm out} = (x_{\rm out}, y_{\rm out}, z_{\rm out})$ and $\mathbf x_{\rm in} = (x_{\rm in}, y_{\rm in}, z_{\rm in})$. The interpolated $\sigma_{vis}$ at $\mathbf x_{\rm in}$ and $\mathbf x_{\rm out}$ are denoted as $\sigma_{vis}^{in}$ and $\sigma_{vis}^{out}$,   Using these values, the surface gradient of $\sigma_{vis}$ at the point $\mathbf x_{12}$ can be computed by, 
\begin{equation}
    (\nabla_s \sigma_{vis})_{p_{12}} = \frac{\sigma_{vis}^{out} - \sigma_{vis}^{in}}{2 dl}.
    \label{probe1}
\end{equation}
Similar procedures are repeated to obtain $(\nabla_s \sigma_{vis})_{p_{23}}$ and $(\nabla_s \sigma_{vis})_{p_{31}}$. Similarly, the surface gradient along vector $\mathbf t$ is obtained, where the interpolation is evaluated at a distance $dl$ on either side of $\mathbf x_{12}$ and along vector $\mathbf t$. This is shown in figure \ref{probing_technique}(b). The interpolated $\sigma_{vis}$ at the probing locations $\mathbf x_{right}$ and $\mathbf x_{left}$ are $\sigma_{vis}^{right}$ and $\sigma_{vis}^{left}$, The surface gradient of $\sigma_{vis}$ along the tangent vector $\mathbf t_{12}$ is then given by, 
\begin{equation}
    (\nabla_s \sigma_{vis})_{t_{12}} = \frac{\sigma_{vis}^{right} - \sigma_{vis}^{left}}{2dl}
    \label{probe2}
\end{equation}

Finally, the distribution of the surface-viscous tension gradient to the Eulerian grid is a straightforward process. The information transfer process is similar to that described for a scalar $X_f$ in Equations \eqref{peskin1}, \eqref{peskin2}, \eqref{peskin3}, and \eqref{peskin4}. Each edge component of the surface-viscous tension gradient (similar to $X_f$) is distributed at the location of the edge centre weighted by one-third of the element area:
\begin{align}
\mathbf{F}_d =&~~~~
\left[ (\nabla_s \sigma_{vis})_{p_{12}} \mathbf{p}_{12} + (\nabla_s \sigma_{vis})_{t_{12}} \mathbf{t}_{12} \right]
\delta(\mathbf{x} - \mathbf{x}_{12}) A_{12} \nonumber \\
&+ \left[ (\nabla_s \sigma_{vis})_{p_{23}} \mathbf{p}_{23} + (\nabla_s \sigma_{vis})_{t_{23}} \mathbf{t}_{23} \right]
\delta(\mathbf{x} - \mathbf{x}_{23}) A_{23} \nonumber \\
&+ \left[ (\nabla_s \sigma_{vis})_{p_{31}} \mathbf{p}_{31} + (\nabla_s \sigma_{vis})_{t_{31}} \mathbf{t}_{31} \right]
\delta(\mathbf{x} - \mathbf{x}_{31}) A_{31}
\end{align}
The last information required to evaluate the dilatational surface viscous forces, $\mathbf F_v^d$, is the evaluation of $\nabla_s \cdot \mathbf u$. We evaluate $\nabla_s \cdot \mathbf u$ on the fly while computing the $\nabla_s \mathbf u$ tensor which is used to evaluate the shear surface viscous forces, $\mathbf F_s$. Hence, we now turn our discussion to the calculation of $\mathbf F_s$, and finally show the on-the-fly evaluation of $\nabla_s \cdot \mathbf u$. 

To evaluate the shear surface viscous forces, $\mathbf F_s$, we use $2\mathbf D_s = \nabla_s \mathbf u \cdot \mathbf I_s + \mathbf I_s \cdot (\nabla_s \mathbf u)^T$, as derived by Scriven \cite{scriven1960dynamics}. Following the formulation of Muradoglu and Trygvasson \cite{muradoglu2008front}, the surface gradient term can be evaluated as a line integral along the edges of an element: 
\begin{align}
\mathbf{F}_s =& \int \int_{A_e} \nabla_s \cdot \left[ \mu_s \left( \nabla_s \mathbf{u} \cdot \mathbf{I}_s + \mathbf{I}_s \cdot (\nabla_s \mathbf{u})^T \right) \right]
~\delta(\mathbf{x} - \mathbf{x}_f)~ dA_e \nonumber \\
=& \int_{C} \left[ \mu_s \left( \nabla_s \mathbf{u} \cdot \mathbf{I}_s + \mathbf{I}_s \cdot (\nabla_s \mathbf{u})^T \right) \right] \cdot \mathbf{p} ~\delta(\mathbf{x} - \mathbf{x}_f)~ dl
\label{shear_discrete}
\end{align}

In this formulation, the main step is to accurately evaluate the surface gradient of the velocity field $\nabla_s \mathbf u$. We utilise our LCRM procedures efficiently to obtain the tensor $\nabla_s \mathbf u$. Because the velocity field is evaluated on an Eulerian grid, $\nabla \mathbf u$ is readily available on the Eulerian grid. The $9$ components of $\nabla \mathbf u$ are interpolated to the mid-point of each edge of the element $e$. For instance, we describe the procedure for evaluating the shear surface viscous forces at the edge \textcircled{1} \textcircled{2}. Suppose the interpolation of $\nabla \mathbf u$ at the midpoint of the edge \textcircled{1} \textcircled{2} is $(\nabla \mathbf u)_{12}$. Leveraging the utility of the level-set distance function in our LCRM formulation, the normal at $\mathbf x_{12}$, that is, $\mathbf n_{12}$ is readily available. Thus, we can evaluate $\nabla_s \mathbf u$ as
\begin{equation}
    (\nabla_s \mathbf u)_{12} = (\nabla\mathbf u)_{12} - \mathbf n_{12}(\mathbf n_{12} (\nabla \mathbf u)_{12}) = \mathbf A_{12}.
\end{equation}
Here, $\mathbf A_{12}$ is a second-order tensor of size $3\times3$. Next, we obtain the surface identity tensor, $\mathbf I_s$ at $\mathbf x_{12}$ as, 
\begin{equation}
    (\mathbf I_s)_{12} = \mathbf I - \mathbf n_{12}\mathbf n_{12} 
\end{equation}
Finally, we evaluate,
\begin{equation}
    (\nabla_s \mathbf u\cdot \mathbf I_s)_{12} = \mathbf A_{12} \cdot (\mathbf I_s)_{12}= \mathbf B_{12}.
\end{equation}
Since $\mathbf I_s$ is idempotent, $(\mathbf I_s \cdot (\nabla_s \mathbf u)^T)_{12} = \mathbf B_{12}^T$. The rate of surface deformation tensor $2\mathbf D_s$ is finally obtained for \textcircled{1} \textcircled{2} as follows:
\begin{equation}
    (2\mathbf D_s)_{12} = (\nabla_s \mathbf u \cdot \mathbf I_s + \mathbf I_s \cdot (\nabla_s \mathbf u)^T)_{12} =  \mathbf B_{12} + \mathbf B_{12}^T
\end{equation}
In a similar fashion, the rate of the surface deformation tensor is evaluated for edges \textcircled{2} \textcircled{3} and \textcircled{3} \textcircled{1}.
Finally, the shear surface viscous force is expressed as
\begin{align}
\mathbf{F}_s &= (\mathbf{F}_s)_{12} + (\mathbf{F}_s)_{23} + (\mathbf{F}_s)_{31} \nonumber \\
&= ~~~\mu_s(\mathbf{x}_{12}) \left( \mathbf{B}_{12} + \mathbf{B}_{12}^T \right) \cdot \mathbf{p}_{12} 
~\delta(\mathbf{x} - \mathbf{x}_{12})~ \Delta s_{12} \nonumber \\
&\quad + \mu_s(\mathbf{x}_{23}) \left( \mathbf{B}_{23} + \mathbf{B}_{23}^T \right) \cdot \mathbf{p}_{23} 
~\delta(\mathbf{x} - \mathbf{x}_{23})~ \Delta s_{23} \nonumber \\
&\quad + \mu_s(\mathbf{x}_{31}) \left( \mathbf{B}_{31} + \mathbf{B}_{31}^T \right) \cdot \mathbf{p}_{31} 
~\delta(\mathbf{x} - \mathbf{x}_{31})~ \Delta s_{31}
\label{fs_discrete}
\end{align}
The shear surface viscosity can be a constant ($a=0$) or a function of the surfactant concentration $(a=1)$. Therefore, $\mu_s(\mathbf x_{12})$ is evaluated as a function of $\Gamma_{12}$ ( discussed in Section \ref{surf_ce}) that is interpolated from the cell centre value, $\Gamma_f$. The distribution of $\mathbf F_s$ to the Eulerian grid is a straightforward process, similar to \eqref{peskin1}, \eqref{peskin2}, \eqref{peskin3}, and \eqref{peskin4}. 

To summarise, the following steps were implemented to evaluate the shear and dilatational surface viscous forces at the interface:
\begin{enumerate}
    \item We evaluate $\nabla \mathbf u$ on the Eulerian grid and the $9$ components of the densor is interpolated to the midpoint of the edges of each element. 
    \item Using the information of $\mathbf n$ on each element, we evaluate the surface gradient of the velocity tensor, $\nabla_s \mathbf u = \nabla \mathbf u - \mathbf n(\mathbf n\cdot \nabla \mathbf u) = \mathbf A$.
    \item The surface divergence of $\mathbf u$ is evaluated by the trace of $\mathbf A$, i.e., $\nabla_s \cdot \mathbf u = \rm tr({\mathbf A})$. This leads to the evaluation of the surface viscous tension, $\sigma_{vis} = (\mu_d - \mu_s)(\rm tr(\mathbf A))$. The evaluation of the normal and tangential forces ($\mathbf F_n, \mathbf F_s$) are similar to the evaluation of $\sigma$, as discussed in this present work as well as in our previous works.
    \item We evaluate the surface identity tensor, $\mathbf I_s =\mathbf I - \mathbf n \mathbf n$ for each edges of the element.
    \item Then we evaluate $\mathbf B = \mathbf A \cdot \mathbf I_s$ and twice of the rate of surface deformation tensor as, $\mathbf B + \mathbf B^T = 2\mathbf D_s$.
    \item Finally, the surface shear viscous forces on the element is obtained by, $\mathbf F_v^s = \sum _k \mu_s (\mathbf x_k) (\mathbf B_k + \mathbf B_k^T) \cdot \mathbf p_k \delta (\mathbf x - \mathbf x_k) \Delta s_k$, where $k$ is the index for the three edges of the element.
\end{enumerate}

\subsection{Surfactant conservation equation} \label{surf_ce}
The surfactant conservation equation on the evolving interface is solved in accordance with the derivation of Muradoglu and Trygvasson, except for the surface diffusion term. A detailed description of the procedure can be found in Shin et al. \cite{shin2018hybrid}. Here, we briefly describe the implementation.
Using Leibniz’s formula, a surface integral of the surface material derivative on an element ($e$) of surface area $A_e$ can be approximated to a change in surfactant mass ($\Gamma A_e$) over a time step $\Delta t$. To compute the diffusion term on the right-hand side, complete information on the Lagrangian interface is required because the surface gradient of $\Gamma$ is significantly dependent on the geometry of the interface. The diffusion term can be computed in a similar manner to implementing the shear surface viscous forces, where the surface gradient term can be evaluated as a line integral along the edges of the element, as described in Equations \eqref{shear_discrete} and \eqref{fs_discrete}. Unlike the computation of $\nabla_s \mathbf u$, $\nabla_s \Gamma$ is computed using the probing technique, as discussed in the evaluation of $\nabla_s \sigma_{vis}$ (equations \eqref{probe1}, \eqref{probe2}). The Lagrangian information of $\Gamma$ is transferred to the Eulerian grid because the message passing of the fields in the Eulerian grid is simpler in parallel processing. This procedure is similar to the method used to transfer the Lagrangian information of scalar $X_f$ to the Eulerian grid (see Equations \eqref{peskin1}, \eqref{peskin2}, \eqref{peskin3}, \eqref{peskin4}). To summarise, the compact formulation of the surfactant transport on the interface is given by, 
\begin{equation}
    \frac{(\Gamma A_e)^{t+\Delta t} - (\Gamma A_e)^t}{\Delta t} = D_\Gamma \left[(\nabla_s \Gamma)_{12}\cdot \mathbf p_{12} \Delta s_{12} + (\nabla_s \Gamma)_{23}\cdot \mathbf p_{23} \Delta s_{23} + (\nabla_s \Gamma)_{31}\cdot \mathbf p_{31} \Delta s_{31} \right],
\end{equation}
which can be rearranged to,
\begin{equation}
    \Gamma^{t+\Delta t} = \Gamma^t A_r + \Delta t \frac{D_\Gamma \left[(\nabla_s \Gamma)_{12}\cdot \mathbf p_{12} \Delta s_{12} + (\nabla_s \Gamma)_{23}\cdot \mathbf p_{23} \Delta s_{23} + (\nabla_s \Gamma)_{31}\cdot \mathbf p_{31} \Delta s_{31} \right]}{A_e ^{t+\Delta t}}.
\end{equation}
Here, $A_r = A_e^{t}/A_e^{t+\Delta t}$ is the area ratio between successive time steps.

Applying Langmuir-Szykowski equation of state for coupling the surface tension, $\sigma$, to the surfactant concentration field, $\Gamma$, we write,
\begin{equation}
    \sigma(\Gamma) = \sigma_0 \left[1 + \frac{\mathcal RT\Gamma_\infty}{\sigma_0} \ln (1 - \Gamma/\Gamma_\infty)\right] = \sigma_0 \left[1 + \beta_s\ln (1 - \Gamma/\Gamma_\infty)\right],
\end{equation}
Because the surface tension coefficient as a function of interfacial concentration is a scalar, as is $\sigma_{vis}$, the normal and tangential forces are computed in a similar fashion as that of $\sigma_{vis}$. 

Our code is validated for surface tension-driven flows with and without surfactants. For a detailed understanding of the various physics underlying the role of surfactants, the reader can refer to \cite{constante2021dynamics, batchvarov2020effect, pico2024surfactant, panda2024marangoni}.

\section{Results and discussion}
In this section, we present the validation of the surface viscous interfacial flows. In Subsection \ref{sec:shear}, we present the validation of the surface shear viscosity by implementing the classical drop deformation test under linear shear flow. In Subsection \ref{sec:rise}, we test a rising drop case to validate the dilatational surface viscosity. In Subsection \ref{sec:wave}, both the surface dilatational and shear viscosities are validated against the parametric surface waves. 
\subsection{Surface-viscous drop under shear flow} \label{sec:shear}
A neutrally buoyant drop in simple linear shear flow is a classical test case for validating surface shear viscosity models. Experiments have shown that the surface shear viscosity is closely linked to the stability of the emulsion system. Flumerfelt \cite{flumerfelt1980effects} utilised a small deformation analysis to incorporate surface viscous effects. Building on this, Phillips et al. \cite{phillips1980experimental} applied this theory to determine the surface viscosity of surfactant-laden drop. Pozrikidis \cite{pozrikidis1994effects} developed a computational model using boundary element methods in the Stokes regime to study the surface viscous effects on a spherical drop; however, his analysis is limited by the high computational cost of calculating surface viscous forces. Gounley et al. \cite{gounley2016influence} also employed boundary element method to evaluate drop deformation and compared them with the Flumerfelt's theory \cite{flumerfelt1980effects}. Luo et al. \cite{luo2019influence} used a finite difference method to study the transient behaviour of drops under shear flow with surface viscosity. 
The primary aim of this study is to demonstrate the accuracy of our level-set-based interface-tracking method in capturing surface viscous effects. 

\begin{figure}[h!]
    \centering
    \includegraphics[width=0.9\linewidth]{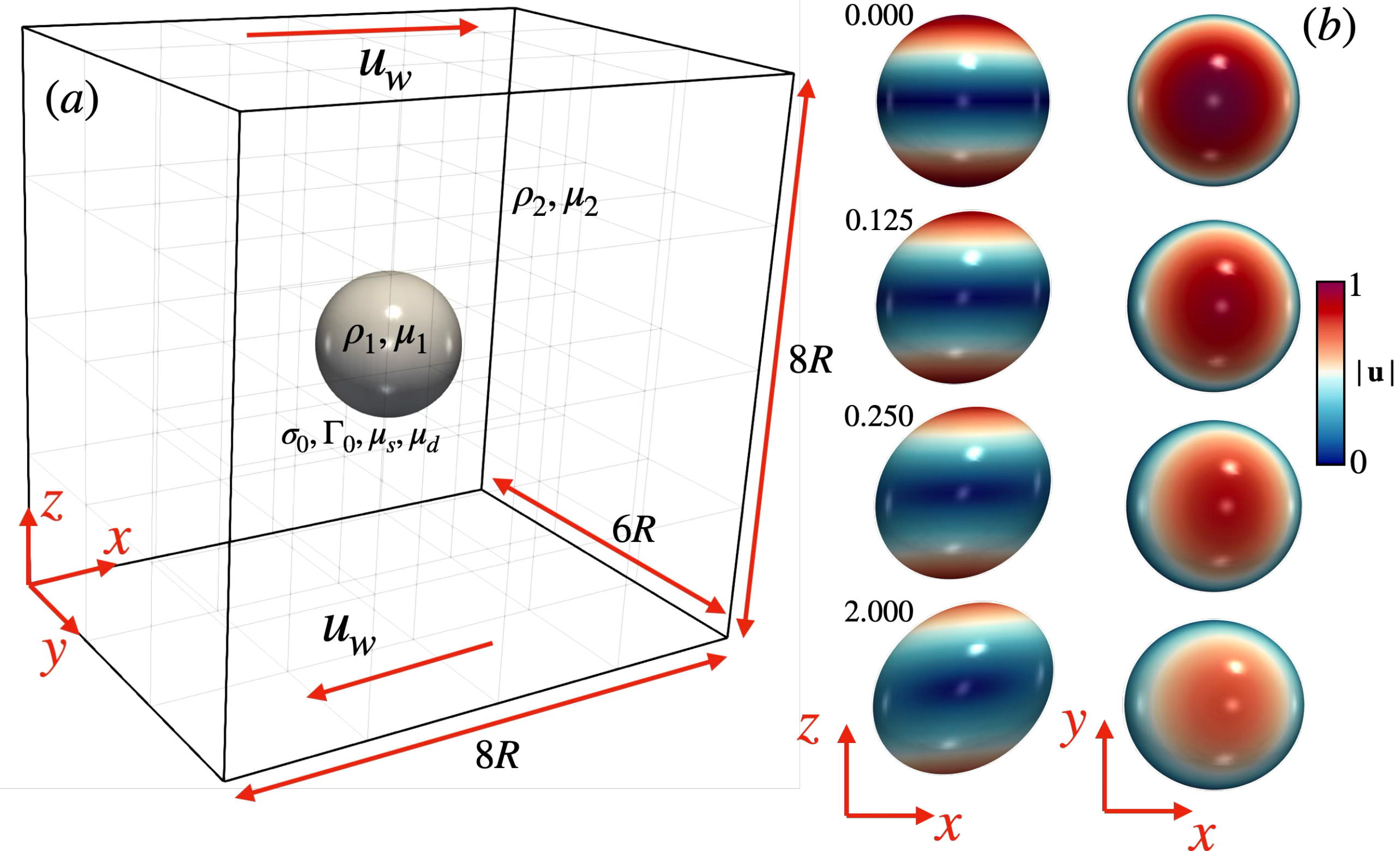}
    \caption{The problem setup for a drop under shear flow is shown in (a) where a drop of density and viscosity $\rho_1$ and $\mu_1$ is under a shear flow due to the continuous motion of the top and bottom boundaries at $u_w$. The background fluid has density $\rho_2$ and viscosity $\mu_2$. The interface separating the two phases has a surface tension $\sigma_0$ in the absence of surfactants. When surfactants are present, the initial coverage is $\Gamma_0$ and the surface dilatational and shear viscosities are $\mu_s$ and $\mu_d$. $R$ is the radius of the drop. The domain is divided into subdomains highlighted by gray cubes inside the domain. The three-dimensional visualisation of the deforming drop is shown in (b) for $Ca = 0.3$ and $Re = 0.1$. Both side ($x-z$) and top ($x-y$) views are shown in (b). The interface is coloured by the magnitude of velocity $|\mathbf u|$ and scaled by $\dot \gamma R$.}
    \label{fig:shear_0}
\end{figure}
The problem setup is illustrated in figure \ref{fig:shear_0}(a). We consider a spherical drop of radius $R$,  density $\rho_1$ and viscosity $\mu_1$ suspended in an ambient liquid of density $\rho_2$ and $\mu_2$. The interface is covered by an insoluble surfactant (such that it is confined to the interface) with an initial surfactant concentration $\Gamma_0$.  The interface exhibits both shear and dilatational surface viscosities, denoted as, $\mu_s$ and $\mu_d$. The clean-interface surface tension is $\sigma_0$. 
The computational domain of interest is a cuboid of dimensions $8R \times 6R \times 8R$, following the configuration used by Luo et al. \cite{luo2019influence}. The domain is decomposed into $48$ subdomains, each containing $32^3$ grid cells. The subdomains are situated as $4\times 3\times 4$, in the $x-$, $y-$, and $z-$ directions, respectively, as shown in figure \ref{fig:shear_0}(a). 
The lateral boundaries are periodic, for both pressure and velocity fields. The top and bottom boundaries move with velocity $u_w$,  generating a linear shear rate, $\dot \gamma = u_w/4R$. Accordingly, the top and bottom boundaries are assigned Dirichlet conditions for the velocity field and Neumann conditions for the pressure field. To ensure that the material derivative of the Heaviside function remains zero, we imposed periodic boundaries laterally and Neumann boundaries at the top and bottom. We assigned the radius of the drop, $R$, and the inverse of the shear rate, $\dot \gamma$, as the length and time scales, respectively, to introduce the dimensionless groups:
\begin{equation}
\begin{split}
    Re &= \frac{\rho_2 \dot \gamma R^2}{\mu_2}, \quad
    Ca = \frac{\mu_2 \dot \gamma R}{\sigma_0}, \quad
    M_\rho = \frac{\rho_2}{\rho_1}, \quad
    M_\mu = \frac{\mu_2}{\mu_1}, \quad
    \beta_s = \frac{\mathcal R T\Gamma_\infty }{\sigma_0}, \\
    Pe &= \frac{\dot \gamma R^2}{\mathcal D}, \quad
    G=\frac{\Gamma_0}{\Gamma_\infty}, \quad
    Bq_s = \frac{\mu_s^\infty}{\mu_2 R}, \quad
    Bq_d = \frac{\mu_d^\infty}{\mu_2 R}.
\end{split}
\end{equation}
Here, $Re, Ca, M_\rho$, and $M_\mu$ are the Reynolds number, Capillary number, density ratio, and viscosity ratio, respectively. The surfactant-related dimensionless numbers are $\beta_s$ (elasticity number), $Pe$ (Péclet number), and G (initial surfactant coverage ratio). $Bq_s$ and $Bq_d$ are the Boussinesq numbers corresponding to the surface shear and dilatational viscosities, respectively. 

\begin{figure}[h!]
    \centering
    \includegraphics[width=0.65\linewidth]{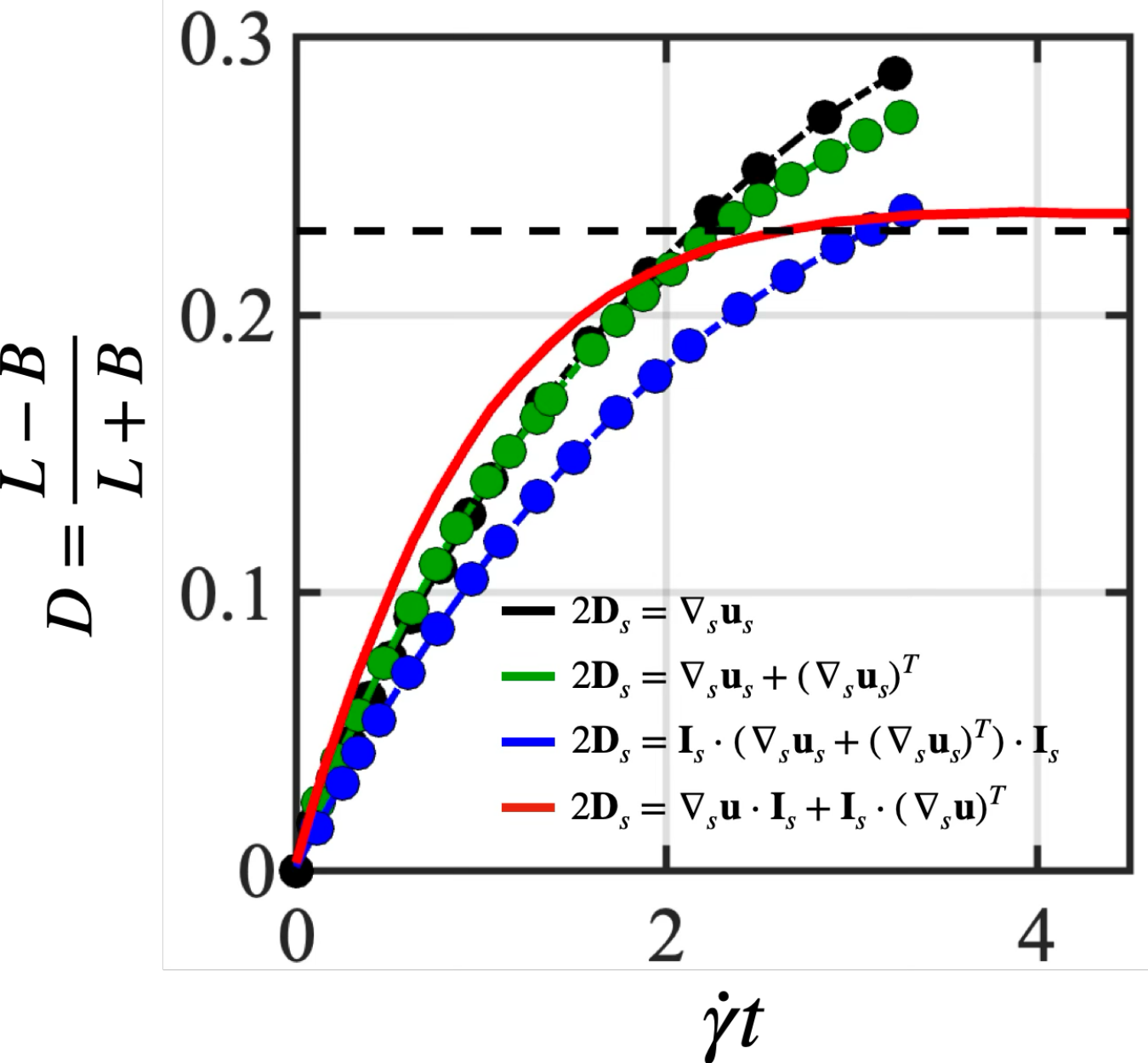}
    \caption{Investigation of different formulation of $2\mathbf D_s$. The parameters are, $Re = 0.05, Ca = 0.33, Bq_s = Bq_d = 5$. The broken line is highlighted for the theoretical prediction by Flumerfelt \cite{flumerfelt1980effects} for $Re \rightarrow 0$.}
    \label{fig:shear_investigation}
\end{figure}

First, we tested the validity of the surface shear viscous forces owing to the surface deformation rate tensor. Thus, we set $Bq_s = Bq_d$ to set $\sigma_{vis} = 0$. Different formulations of the surface deformation rate tensor are evaluated in figure \ref{fig:shear_investigation}. Here, $\mathbf u_s$ represents the surface projection of the interfacial velocity, $\mathbf u_s = \mathbf u \cdot \mathbf I_s$. We found that an accurate representation of the surface deformation rate tensor is significant in evaluating surface shear viscous forces. Not only does our choice of $2\mathbf D_s$ differ significantly from other formulations in the transient state, it also converges to the theoretical prediction by Flumerfelt \ref{fig:shear_investigation}. We will later show that our choice of $2\mathbf D_s$ also captures the transient state by comparing it with Pozrikidis \cite{pozrikidis1994effects}. This confirms that the surface gradient operates on the interfacial velocity, $\mathbf u(\mathbf x = \mathbf x_f)$ and not on the surface projected interfacial velocity, $\mathbf u_s$, as used by Lopez and Hirsa \cite{lopez1998direct}. The surface gradient on the surface velocity results in additional terms owing to the presence of $\nabla_s \mathbf I_s$ \ref{appA}. This results in the surface projection of a third-rank tensor which is a product of the curvature tensor, $\mathbf K = \nabla \mathbf n$ and the normal vector, $\mathbf n$. In the case of Lopez and Hirsa \cite{lopez1998direct}, this term has negligible effects owing to the linearisation of the problem, where the base state is a flat interface. In such cases, where the principal curvature is zero, it is evident that this additional term does not affect the final outcome of their work. However, the additional term cannot be neglected in three-dimensional generalised flows; therefore, a correct formulation of $2\mathbf D_s$ is significant.        

Next, the parameters are set to $Re = 0.1, Ca = 0.1,$ and $M_\rho = M_\mu = 1$. To ensure that the surface tension remains unaffected by surfactants, we fix $\beta_s = 0$, while $G = 0.5$ and $Pe = 100$. Under these conditions, surfactant transport behaves purely as a passive surface-scalar transport process at the interface. Following Luo et al. \cite{luo2019influence}, the choice of $Re = 0.1$ allows meaningful comparison with existing literature, where most of studies are conducted in the Stokes-flow regime. The transient deformation of the drop for the clean interface is shown in figure \ref{fig:shear_0}(b). Starting from an initial spherical shape, the drop deformed under the imposed shear flow, reaching its maximum deformation. Three-dimensional visualisation from the side ($xz$) and top ($xy$) at $\dot \gamma t = 0.0, 0.125, 0.25, $ and $2.0$ are shown in figure \ref{fig:shear_0}(b), where the interfacial velocity is coloured on the drop interface.

\begin{figure}[h!]
    \centering
    \includegraphics[width=0.9\linewidth]{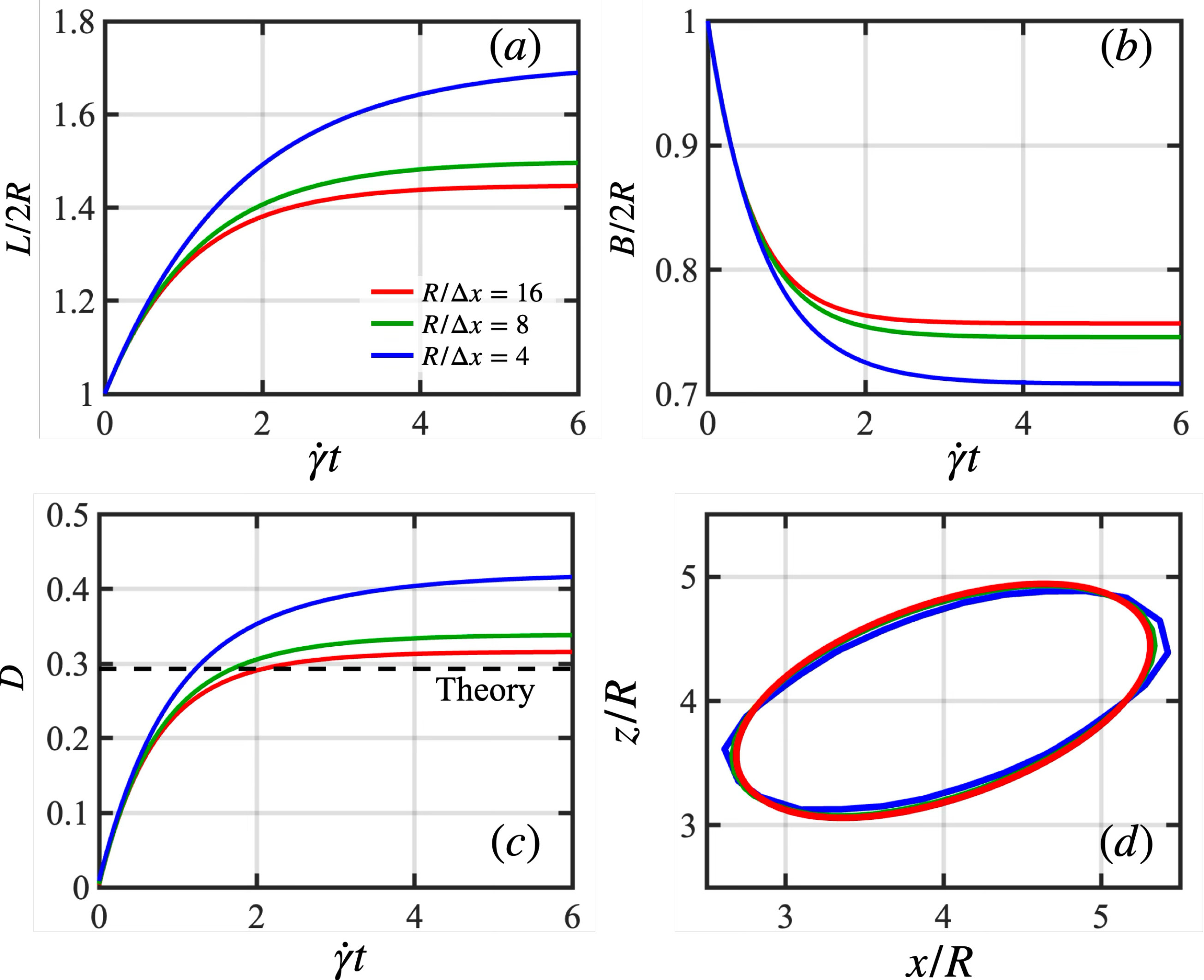}
    \caption{Mesh grid independence test for $Ca = 0.3$ and $Bq_s = 0.5$: Temporal evolution of the (a) semi-major axis ($L$), (b) semi-minor axis ($B$), (c) deformation parameter, $D = (L-B)/(L+B)$, and (d) interfacial contour at $\dot \gamma t = 3.00$ for $R/\Delta x = 16, 8, $ and $4$. The black dotted line in (c) highlights the theoretical prediction by Flumerfelt \cite{flumerfelt1980effects} for $Re \rightarrow 0$.}
    \label{fig:shear_1}
\end{figure}

A mesh independence test is performed for $Bq_s = 0.5$, $Bq_d = 0.0$ and $Ca = 0.3$. The results are shown in figure \ref{fig:shear_1}. The major and minor axes of the deforming drop are denoted as $L$ and $B$. The test is carried out for three grid sizes, that is, $R/\Delta x = 4, 8,$ and $16$. The temporal evolution of $L$ and $B$ is shown in figure \ref{fig:shear_1}(a,b). As the resolution increased, the deformation ($L$ and $B$) converged to a constant value. In figure \ref{fig:shear_1}(c), we evaluated the deformation parameter $D = (L-B)/(L+B)$ and compared it with the theoretical prediction \cite{flumerfelt1980effects}. As the resolution increases, the accuracy of evaluating $D$ approaches the theoretical prediction. Thus, this test not only guarantees mesh grid convergence but also illustrates the accuracy of our level-set-based front-tracking method. A comparison of the deformed drops with different mesh resolutions at $\dot \gamma t = 3$ is shown in figure \ref{fig:shear_1}(d).   

\begin{figure}[h!]
    \centering
    \includegraphics[width=1.0\linewidth]{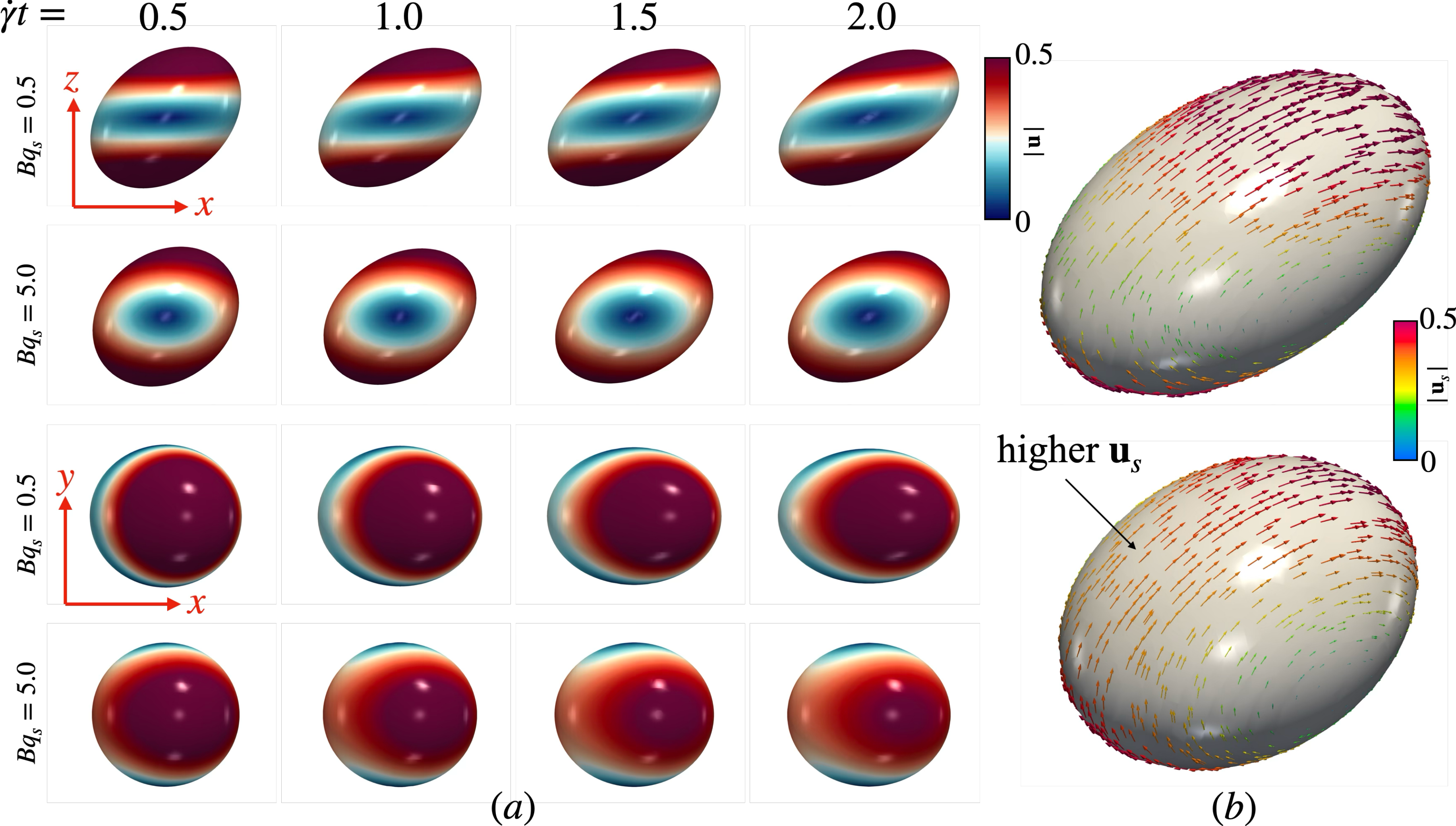}
    \caption{(a)Three-dimensional visualisation of the deforming drops under shear flow for $Ca = 0.3$ and $Bq_s = 0.5$ and $5.0$. The drop is coloured by the magnitude of the velocity at the interface ($|\mathbf u|$). The surface velocity ($\mathbf{u}_s$) quivers on the deformed drop is shown in (b) for $Bq_s = 0.5$ (top) and $5.0$ (bottom) at $\dot \gamma t = 2.0$.}
    \label{fig:shear_2}
\end{figure}

Three-dimensional visualisation for two cases where $Ca = 0.3, Re = 0.1, Bq_d = 0.0$, but $Bq_s = 0.5$ and $5.0$ are shown in figure \ref{fig:shear_2} to indicate the importance of the shear surface viscosity. In figure \ref{fig:shear_2}(a), the temporal evolution of the drop for $Bq_s = 0.5$ and $5.0$ is shown at a period of $\dot \gamma t = 0.5$. The first two panels compare the interface profile in the $xz$ plane, and the bottom two are for the $xy$ plane. The drop interface is coloured by the magnitude of the interfacial velocity, $\mathbf u$. A stark difference is observed not only in the deformation of the drop but also in the interfacial velocity across the surface. At lower $Bq_s$, the drop is deformed, and the interfacial velocity is highly dependent on the background linear shear flow as $|\mathbf u|$ is the lowest at $z = 4R$ (the centre of $z-$ plane) of the domain. However, at higher $Bq_s$, the surface shear viscosity tends to lower the velocity gradient at the interface. The surface flow becomes almost circular in the $xy$ plane, as shown in the second panel of figure \ref{fig:shear_2}(a). This is also evident from the top view, where the surface velocity gradient is slow for $Bq_s = 5$ as opposed to $Bq_s = 0.5$. The surface velocity gradient is reduced owing to the replenishment of the surface flow at the dilated region of the deforming drop. This is illustrated in figure \ref{fig:shear_2} where glyphs of higher surface velocity, $\mathbf u_s$, are observed at the dilated zone of the interface for $Bq_s = 5.0$ compared to $Bq_s = 0.5$. These qualitative comparisons are in good agreement with those of Gounley et al. \cite{gounley2016influence}.

\begin{figure}[h!]
    \centering
    \includegraphics[width=0.75\linewidth]{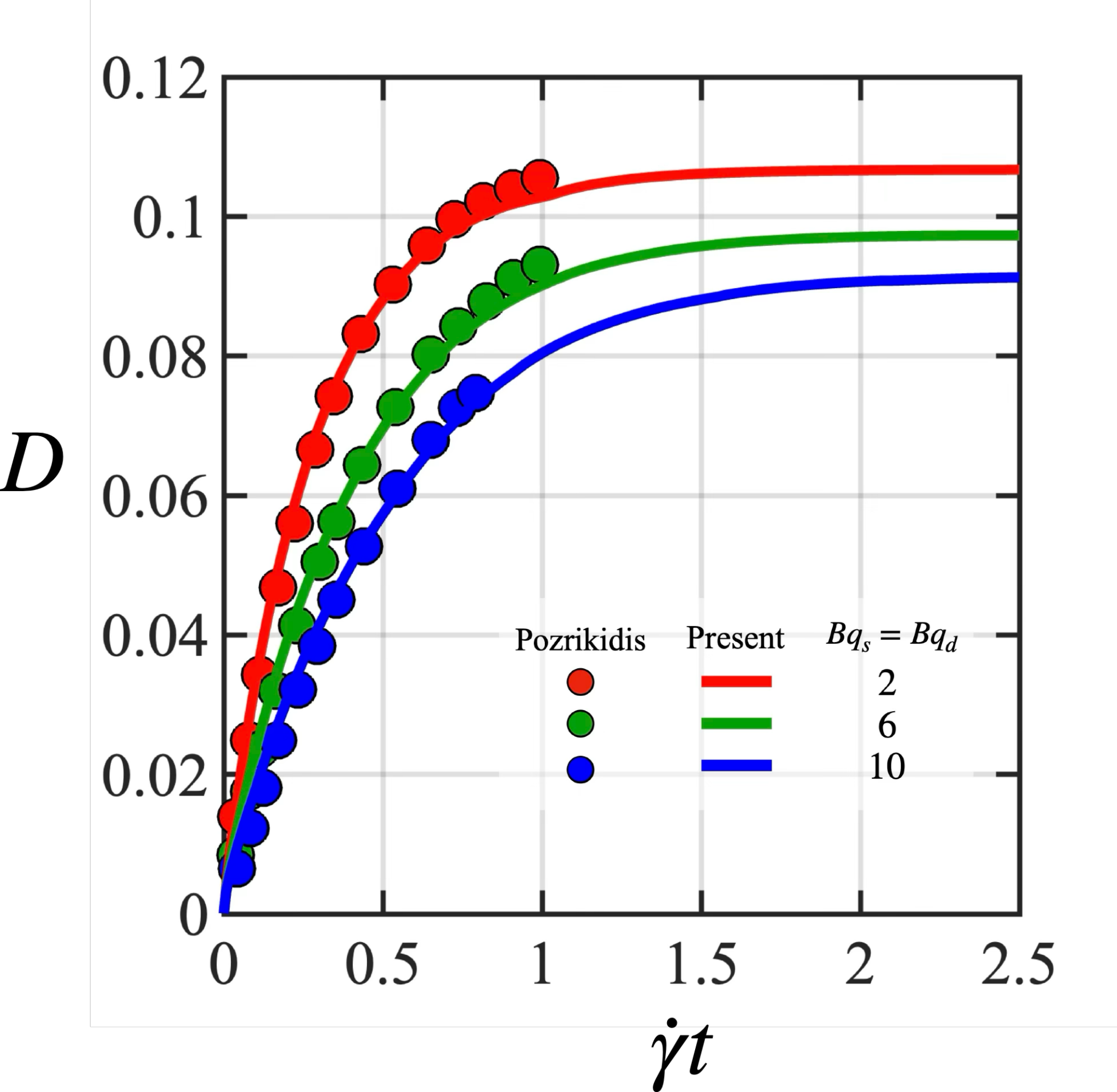}
    \caption{Temporal evolution of the deformation parameter, $D$ for $Bq_s=Bq_d = 0, 2, 6, 10$ is compared with Pozrikidis \citep{pozrikidis1994effects}.}
    \label{fig:shear_3}
\end{figure}

A quantitative validation is presented in Figure \ref{fig:shear_3}, where we evaluate the deformation parameter, $D = (L-B)/(L+B)$. In all cases, we set $Bq_s = Bq_d$ to eliminate the influence of surface compressibility ($\nabla_s \cdot \mathbf u$), thus isolating the effect of surface shear viscosity by responding to surface deformation ($2\mathbf D_s$). The Boussinesq number is varied over 0, 2, 6, and 10, and the transient evolution of $D$ is compared with the boundary element method results of Pozrikidis \cite{pozrikidis1994effects}. 
Our simulations showed excellent agreement with Pozrikidis' results. Increasing $Bq_s$ slows the deformation rate and lowers the steady-state deformation, reflecting the enhanced resistance of the interface to tangential flow when the surface viscosity is higher. Unlike the boundary element method, our LCRM allows the simulation to be extended reliably to steady state, demonstrating both the accuracy and robustness of the method for capturing interfacial viscous effects. It should be noted that $D$ for the clean case is similar to the predictions of Luo et al. \cite{luo2019influence} compared to the computational results of Pozrikidis \cite{pozrikidis1994effects}.

\begin{figure}[h!]
    \centering
    \includegraphics[width=0.8\linewidth]{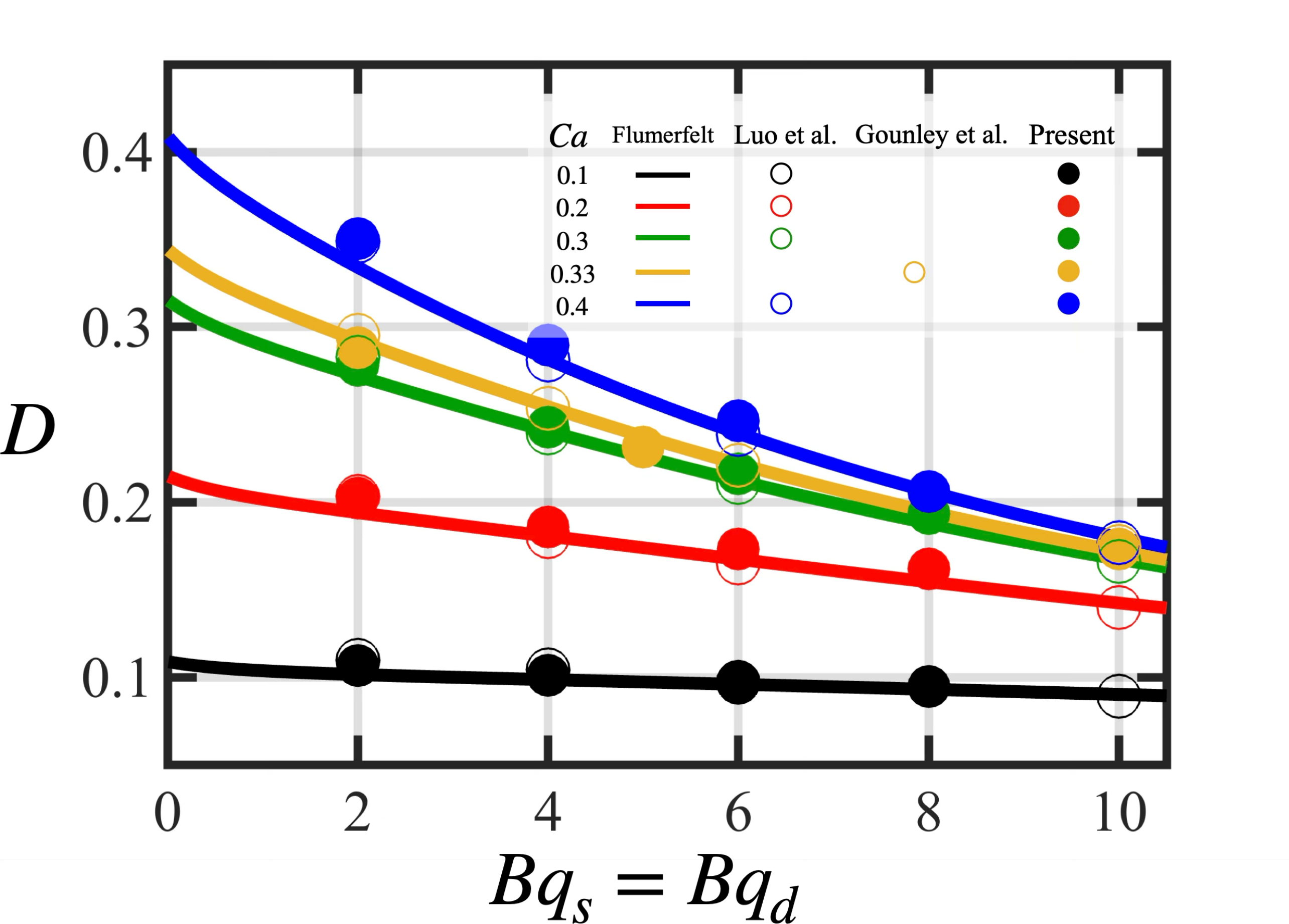}
    \caption{\textbf{Comparison of steady-state deformation parameter with the analytical solution of Flumerfelt \citep{flumerfelt1980effects}, boundary element method of Gounley et al. \cite{gounley2016influence} and finite-difference method of Luo et al. \citep{luo2019influence}.}}
    \label{fig:shear_4}
\end{figure}

In figure \ref{fig:shear_4}, we evaluated the steady-state deformation parameter for varying capillary and Boussinesq numbers. All simulations are run for $\dot \gamma t = 5$. The evaluated $D$ are then compared with the explicit expression of Flumerfelt and the computational results of Luo et al. \cite{luo2019influence} and Gounley et al. \cite{gounley2016influence}. By increasing $Ca$, the variation in $D$ is significant and a similar trend is captured by our level-set-based interface tracking method. Our results are observed to be overestimated but within $4\%$ of the error from Flumerfelt's theory. The overestimation can be a consequence of the finite $Re$ used in this study. The sensitivity of inertial effects to the shearing drop can be tested by reducing $Re$. However, we are focused on showing the validity of our method in the presence of surface viscous effects.

\subsection{Rising surface viscous drop in a quiescent fluid}
\label{sec:rise}
\begin{figure}[h!]
    \centering
    \includegraphics[width=0.7\linewidth]{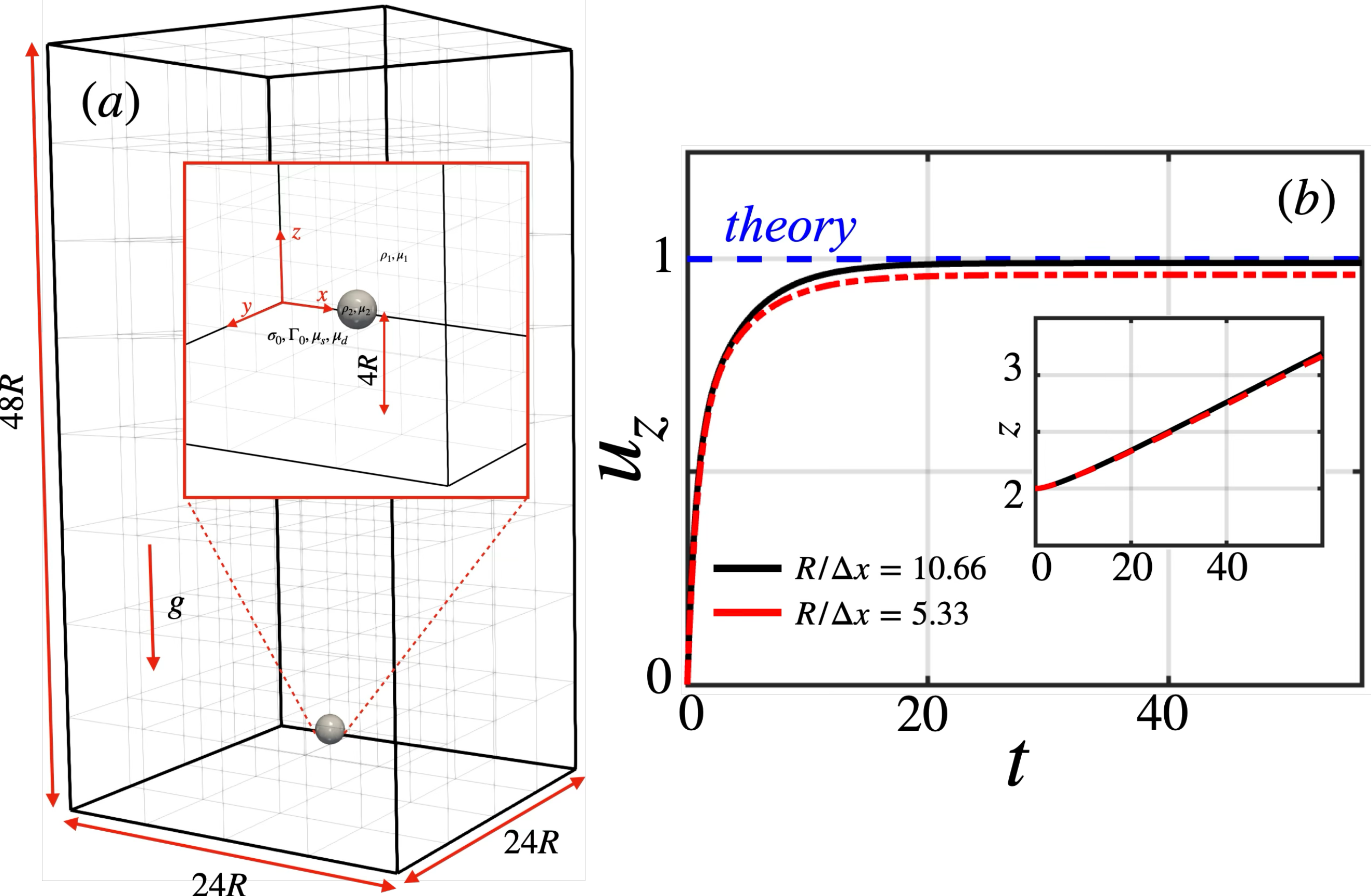}
    \caption{Problem statement is summarised in (a) where a drop is initially at rest and is allowed to rise due to gravity, $g$. The nomenclature of various physical properties related to the drop's interface is shown in the inset of (a), where it is also shown that the drop is at a height of $4R$ at $t=0$. The temporal evolution of the velocity and position of the drop is shown in (b) for two different mesh sizes and compared with the theory. Here, the velocity is scaled by the clean drop rise, $V_c = 0.006538 ~ \rm{m/s}$. The length and time are scaled by $R$ and $\sqrt{R/g}$, respectively.}
    \label{fig_res_0}
\end{figure}
Drop migration in stagnant fluids is a classical problem in fluid mechanics. Lebedev \cite{lebedev1916stokes} and Silvey \cite{silvey1916fall} observed that contaminated drops and bubbles migrate under gravity in a manner similar to a solid sphere. This behaviour directly implies that the presence of interfacial agents significantly alters the tangential boundary condition at the interface. Boussinesq \cite{boussinesq1913existence} formalised the concept of surface viscosity to explain such phenomena. Edwards and Wasan \cite{wasan1992interfacial}, following the derivation of Levan \cite{levan1981motion}, obtained an expression for the migration velocity as a function of surface viscosity and demonstrated that, in the Stokes regime, the migration velocity is independent of the surface shear viscosity. Narsimhan \cite{narsimhan2018effect} elucidated the underlying mechanism and concluded that the migration velocity of a surface-viscous drop could be determined using a modified effective viscosity. Dehghani and Narsimhan \cite{dehghani2019unsteady} evaluated the drag on viscoelastic drops in the unsteady Stokes regime. Similarly, Reusken et al. \cite{reusken2013numerical} and Dandekar et al. \cite{dandekar2020effect} studied the influence of surface viscosity on drops in the Poiseuille flow, while Singh and Narsimhan \cite{singh2021impact} investigated its effects on initially prolate and oblate drops. Here, we intend to show the validation of the dilatational surface viscosity.

\begin{figure}[h!]
    \centering
    \includegraphics[width=1.0\linewidth]{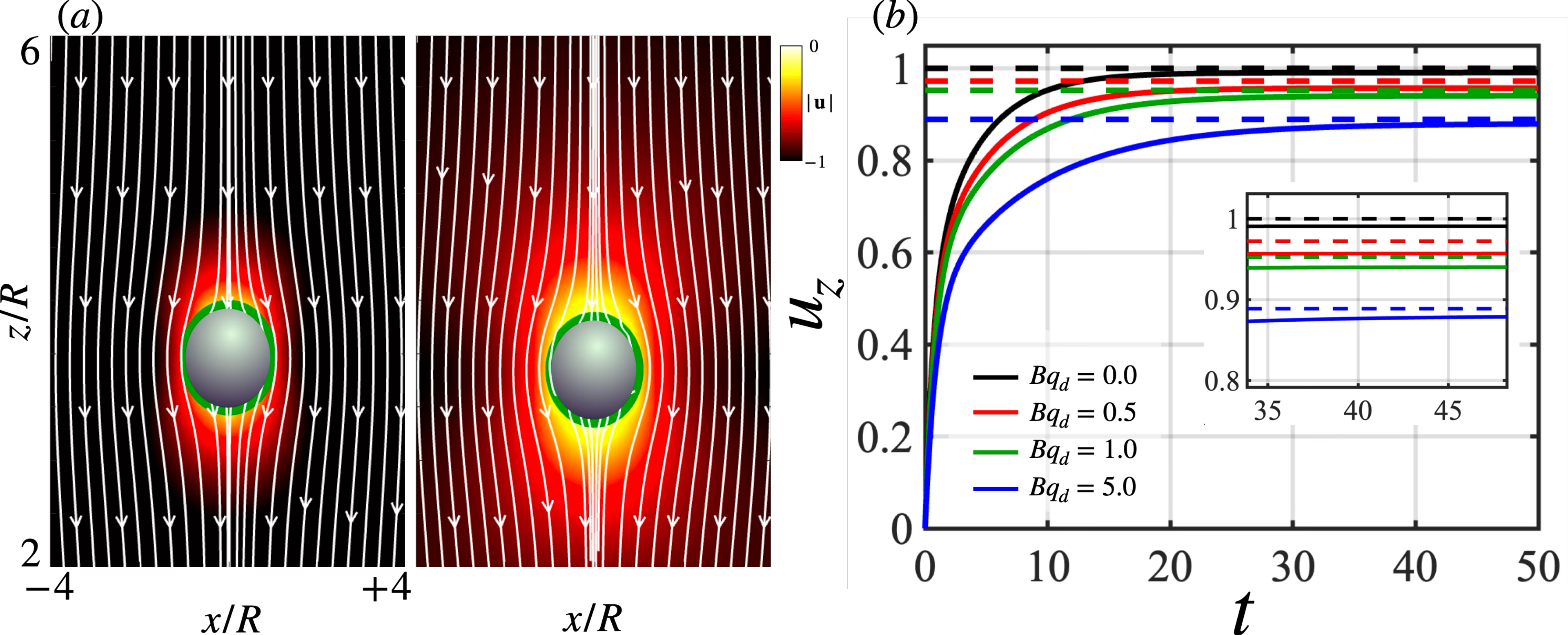}
    \caption{The $z-$ velocity of the drop rise for $Bq_d = 0, 0.5, 1, 5$ is shown in (a) where a dotted line is highlighted to signify the theoretical prediction for the clean case. A parametric study is shown in (b) for varying $Bq_d$ and is compared with the theoretical prediction. The grid convergence test is shown in (c) for $Bq_d = 0, 0.5, 1,$ and $5$, for which first-order accuracy is attained, as highlighted by the black dotted line. }
    \label{fig_res_1}
\end{figure}
The computational domain of interest is shown in figure \ref{fig_res_0}(a), where a buoyant drop of radius $R$ is initially placed at a distance of $4R$ from the bottom face (see the inset of figure \ref{fig_res_0}(a)). The drop has density $\rho_2$ and viscosity $\mu_2$ and is submerged in a stagnant fluid with density $\rho_1$ and viscosity $\mu_1$. The contaminated surface has an initial surfactant concentration $\Gamma_0$ and shear and dilatational surface viscosities $\mu_s$ and $\mu_d$. The surface tension of the same drop but with a clean interface is $\sigma_0$.  
The computational domain is of size $24R \times 24R \times 48R$ to avoid boundary effects in determining the terminal velocity of the drop. The lateral boundaries are periodic, and the top and bottom boundaries are free-slip for the velocity fields. For the pressure field, the lateral boundaries are periodic, and the top and bottom boundaries are Neumann. 

The drop radius, $R$, and the freefall velocity, $\sqrt {gR}$, are chosen as the length and velocity scale, respectively, in the problem. The hydrodynamic dimensionless numbers involved in the problem are 
\begin{equation}
    Re = \frac{|\rho_1-\rho_2| \sqrt{gR^3}}{\mu_1}, Bo = \frac{|\rho_1-\rho_2| g R^2}{\sigma_0}, M_\rho = \frac{\rho_2}{\rho_1}, M_\mu = \frac{\mu_2}{\mu_1}.
\end{equation}
where $Re$ and $Bo$ are the Reynolds and Bond numbers, respectively. The surface viscous ($Bq_s, Bq_d$) and surfactant dynamics parameters ($\beta_s, Pe, G$) remain the same as those discussed in the previous section. The migration terminal velocity, $U_{\rm mig}$, is then given by, 
\begin{equation}
    \frac{U_{\rm mig}}{\sqrt{gR}} = \frac{2}{9}~Re ~\left(1 + \frac{1}{2}\left(1 + Bq_d + \frac{3}{2}M_\mu\right)^{-1}\right). 
    \label{migration}
\end{equation}

\begin{table}[h!]
    \centering
    \footnotesize
    \renewcommand{\arraystretch}{1.3} 
    \setlength{\extrarowheight}{2pt}  

    \caption{Comparison of the terminal velocity for varying $Bq_d$ with the theoretical prediction. Here, the superscripts ``coarse'' and ``fine'' refer to the mesh sizes $R/\Delta x = 5.33$ and $10.66$, respectively. }
    \label{tab_3}

    \begin{tabular}{p{0.1\textwidth}p{0.1\textwidth}p{0.1\textwidth}p{0.1\textwidth}p{0.1\textwidth}p{0.1\textwidth}} 
        \hline
        \textbf{$Bq_d$} & ~$U_{\rm mig}$ & ~$U_{\rm {DNS}}^{\text{coarse}}$ &~~$U_{\rm {DNS}}^{\text{fine}}$ & $\Delta_{\rm {coarse}}(\%)$ & $\Delta_{\rm {fine}}(\%)$ \\[3pt]
        \hline
        0.0 & 0.006538 & 0.006297 & 0.006477 & ~~3.68 & ~0.93\\
        0.5 & 0.006495 & 0.006059 & 0.006256 & ~~6.71 & ~3.68\\
        1.0 & 0.006356 & 0.005915 & 0.006147 & ~~6.94 & ~3.29\\
        5.0 & 0.005811 & 0.005327 & 0.005749 & ~~8.33 & ~3.30\\
        \hline
    \end{tabular}

\end{table}

We set $Re = 0.35$ to run the cases in the Stokes regime. The Bond number is set to $7.66\times 10^{-2}$ to ensure that the drop remains spherical. The density ratio $M_\rho$ is fixed at $0.9$, and the viscosity ratio is maintained at $M_\mu = 1$. The elasticity number $\beta_s$ is equal to $0$, to decouple the effects of surface viscosity from elasticity. Moreover, $Bq_s = 0$ was set to focus solely on the validation of the dilatational surface viscosity. 

We tracked the position of the drop in time and obtained the vertical velocity as a function of time. An example is shown in figure \ref{fig_res_0}(b). As shown in the inset of figure \ref{fig_res_0}(b), after a certain period of time, the drop rises linearly and thus attains a terminal migration velocity. This is shown for two different mesh sizes in the clean case. Our code agrees well with the theoretical prediction, as highlighted by the blue dotted line in Figure \ref{fig_res_0}(b). 

To assess the effects of the dilatational surface viscosity, we ran three cases for $Bq_d = 0.5, 1, $ and $5$ and for two different mesh sizes, $R/\Delta x = 5.33$ and $ 10.66$. To compare the profiles of the drop, we present the velocity contours in the frame of reference of the rising drop for $Bq_d = 0$ and $1$ in figure \ref{fig_res_1}(a). The streamlines in the frame of reference of the rising drop are also overlaid on the velocity contours. The streamlines in both cases are parallel to the $z-$axis, except near the interface. This confirms that the drop is in the Stokes regime. The velocity contours signify that the momentum is more diffusive for $Bq_d = 1$. This results in a deceleration of the rising drop in the presence of surface viscous effects. Therefore, the position of the drop for the surface viscous case is lower than that for the clean case. 

The temporal evolution of the vertical velocity for these cases is shown in \ref{fig_res_1}(b). The dotted lines of the respective colour codes represent the theoretical prediction of $U_{\rm mig}$. The vertical velocity is scaled by $U_{\rm mig}$ for the clean case. The retardation of the rising drop in the case of $Bq_d = 5$ is significant compared to that of $Bq_d = 0.5$ and $1$. The simulations are carried out until the drop attained a steady state of rising velocity. Qualitatively comparing the velocity at $t = 50$, our DNS results agree well with the theoretical prediction. 

As shown in the inset of figure \ref{fig_res_1}(b), the terminal velocity is underestimated in both the clean and surface viscous cases. A grid dependence test is carried out for coarser ($R/\Delta x = 5.33$) and finer ($R/\Delta x = 10.66$) in table \ref{tab_3}. In both cases, the estimated terminal velocities are underestimated compared to $U_{\rm mig}$. However, increasing the resolution results in at least $3\%$ error ($\Delta(\%) =100\times (U_{\rm mig}-U_{\rm DNS})/U_{\rm mig}$). Therefore, increasing the resolution can increase the accuracy of the numerical method.          
\subsection{Parametric waves on a surface viscous interface}
\label{sec:wave}
\begin{figure}[h!]
    \centering
    \includegraphics[width=1.0\linewidth]{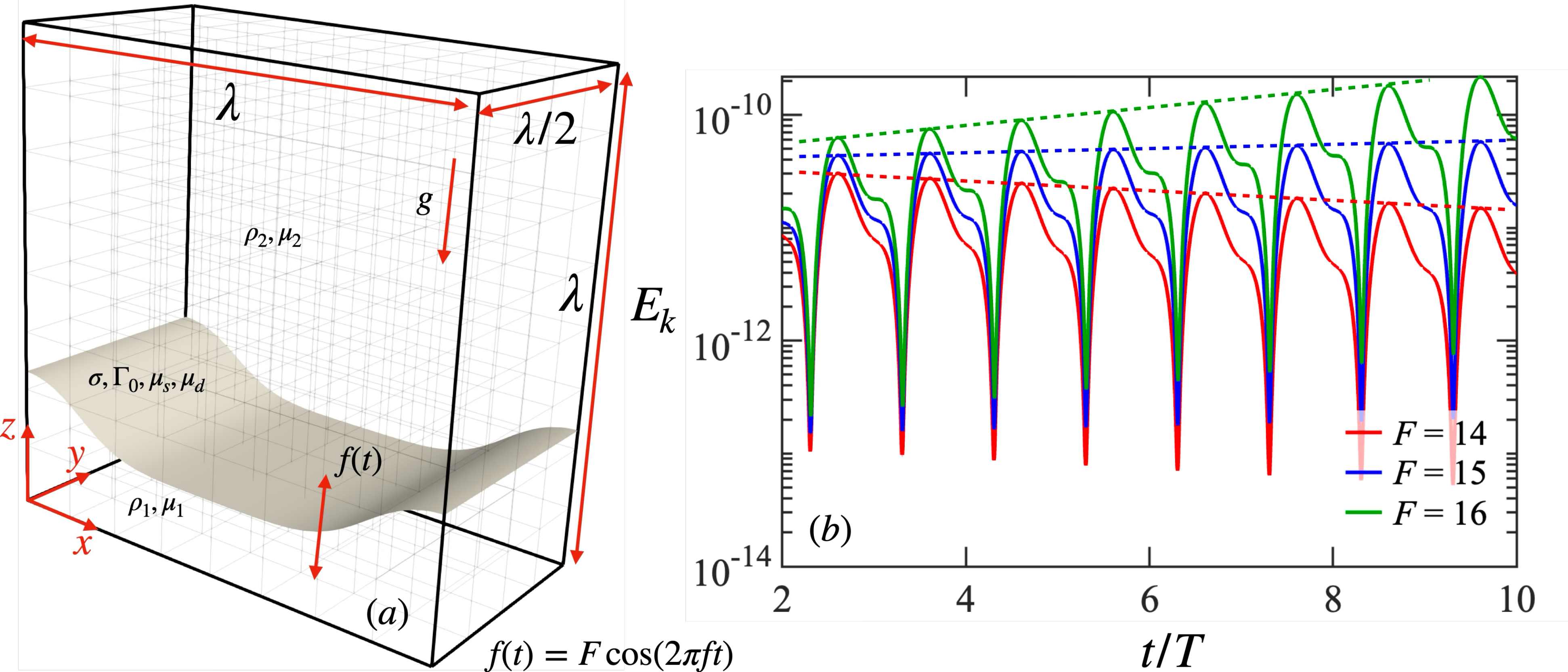}
    \caption{The problem setup is shown in (a) where an interface coloured gray is under a parametric vibration $f(t) = F \cos (2\pi f t)$. The computational domain is a cuboid of size $\lambda \times \lambda/2 \times \lambda$ and the domain is subdivided into $8\times 4 \times 8$ subdomains of $16^3$ grids each. The temporal evolution of the kinetic energy is shown in (b), where the interface is surface viscous with $Bq = 1$ and $\beta_s = 0$. The dotted line highlights the slope at which the energy is either growing or decaying.}
    \label{fig:far_0}
\end{figure}

The third case of surface viscosity validation is tested for parametric surface waves. When a fluid interface is vibrated at a certain amplitude and frequency, interfacial waves are observed, which usually oscillate at a frequency twice that of the forced vibration \cite{miles1990parametrically}. Although surface waves damping has been well studied in the past, parametric surface waves have been studied for applications in pattern formation \cite{panda2024marangoni}, quantum hydrodynamics \cite{bush2020hydrodynamic}, and atomisation \cite{james2003vibration}. Faraday waves with surfactants have also been studied to measure the damping effects of surfactants \cite{henderson1998effects} and for pattern transitions on the surface \cite{panda2024marangoni}. However, these studies typically addressed Marangoni-driven flows because of concentration gradients. Ubal et al. \cite{ubal2005influence} investigated the role of surface viscosity in the excitation of parametric surface waves. Following the numerical experiment of Ubal et al., \cite{ubal2005influence}, we first aimed to validate the threshold acceleration at which the surface waves grew over time. The problem setup is illustrated in figure \ref{fig:far_0}(a), where a cuboid encompasses two fluids with densities $\rho_1, \rho_2$ and viscosities $\mu_1, \mu_2$. The two phases are distinguished by the interface of surface tension $\sigma_0$ (for the clean case), initial surfactant coverage $\Gamma_0$ and dilatational and shear surface viscosities $\mu_d$ and $\mu_s$. The wavelength of the domain is $\lambda$. We chose a size domain $\lambda \times \lambda/2 \times \lambda$. Initially, the interface is flat and at a height $h \ll \lambda$. The top and bottom boundaries are Dirichlet for velocity, such that they satisfy the no-slip and no-penetration boundary conditions. Periodic boundaries are imposed on the lateral sides. An external sinusoidal volumetric force is applied at a frequency $f$ and acceleration amplitude $A$. Choosing $\lambda$ as the length scale and the inverse of angular frequency $\Omega = 2\pi f$ as the time scale, the dimensionless groups utilised are 
\begin{align}
    F = \frac{A}{g}, ~~Re = \frac{\rho_1 \Omega \lambda^2}{\mu_1}, ~~ We = \frac{\rho_1 \Omega^2 h^3}{\sigma_0}, ~~ M_\rho = \frac{\rho_2}{\rho_1}, ~~ M_\mu = \frac{\mu_2}{\mu_1},\nonumber \\ ~~ G = \frac{\Gamma_0}{\Gamma_\infty}, \beta_s = \frac{RT\Gamma_\infty}{\sigma_0},  Pe = \frac{\lambda^2\Omega}{\mathcal D}, ~ Bq_s = \frac{\mu_s^\infty}{\mu_1 \lambda}, ~~ Bq_d = \frac{\mu_d^\infty6}{\mu_1 \lambda}.
\end{align}
According to Ubal et al. \cite{ubal2005influence}, the initial height is $h = 10^{-3}~ \rm m$ and $\lambda = 4.986\times10^{-3}~\rm m$. The density and viscosity ratios are $M_\rho = 10^{-3}$ and $M_\mu = 10^{-2}$, respectively. The frequency of vibration is set at $100 ~\rm Hz$, that is, $\Omega = 200\pi ~\rm rad~s^{-1}$. The Reynolds and Weber numbers are $624.8$ and $889.72$. Because $Re \gg 1$, this case also signifies the utility of our code in the unsteady regime. The surfactant properties are chosen such that $G = 0.5$ and $Pe = 100$. These parameters are fixed in our study. The parameters to be varied are the surface dilatational and shear Boussinesq numbers ($Bq_s, Bq_d$) and $a$. 

First, we set the elasticity number $\beta_s = 0$ and $a = 0$ to decouple the effects of elasticity. We followed a procedure similar to that prescribed by Perinet et al. \cite{perinet2009numerical} to assess the threshold acceleration at which the interface becomes unstable to external vibrations. This method has been proven to be robust, as it can be readily extended to surfactant-covered interfaces \cite{panda2024marangoni}. First, we choose certain acceleration amplitudes at which the interface is under vibration, say $F = 14, 15,$ and $16$, as shown in figure \ref{fig:far_0}(b). We evaluated the total kinetic energy of the system upon vibration. Because the vibration is periodic at a frequency $f$, or time period $T$, the kinetic energy response is also oscillatory and of the same time period $T$. One of the best ways to assess the instability of an interface is to track whether $E_k$ grows or decays with time. The slope of $E_k$ over time for $F=14,15,$ and $16$ is the growth of the system, where the slope is $<0$ for $F=14$, and $>0$ for $F=16$. At $F=15$, the slope is $\approx 0$, indicating that the system is at the onset of instability. To quantify the threshold acceleration, we imposed a linear interpolation of the growth to the acceleration amplitude. This is true when the driving parameter $F$ is close to the pitchfork bifurcation. For the case of $Bq_d =1, Bq_s = 0$, the threshold acceleration obtained is $14.77$.  

\begin{figure}[h!]
    \centering
    \includegraphics[width=0.75\linewidth]{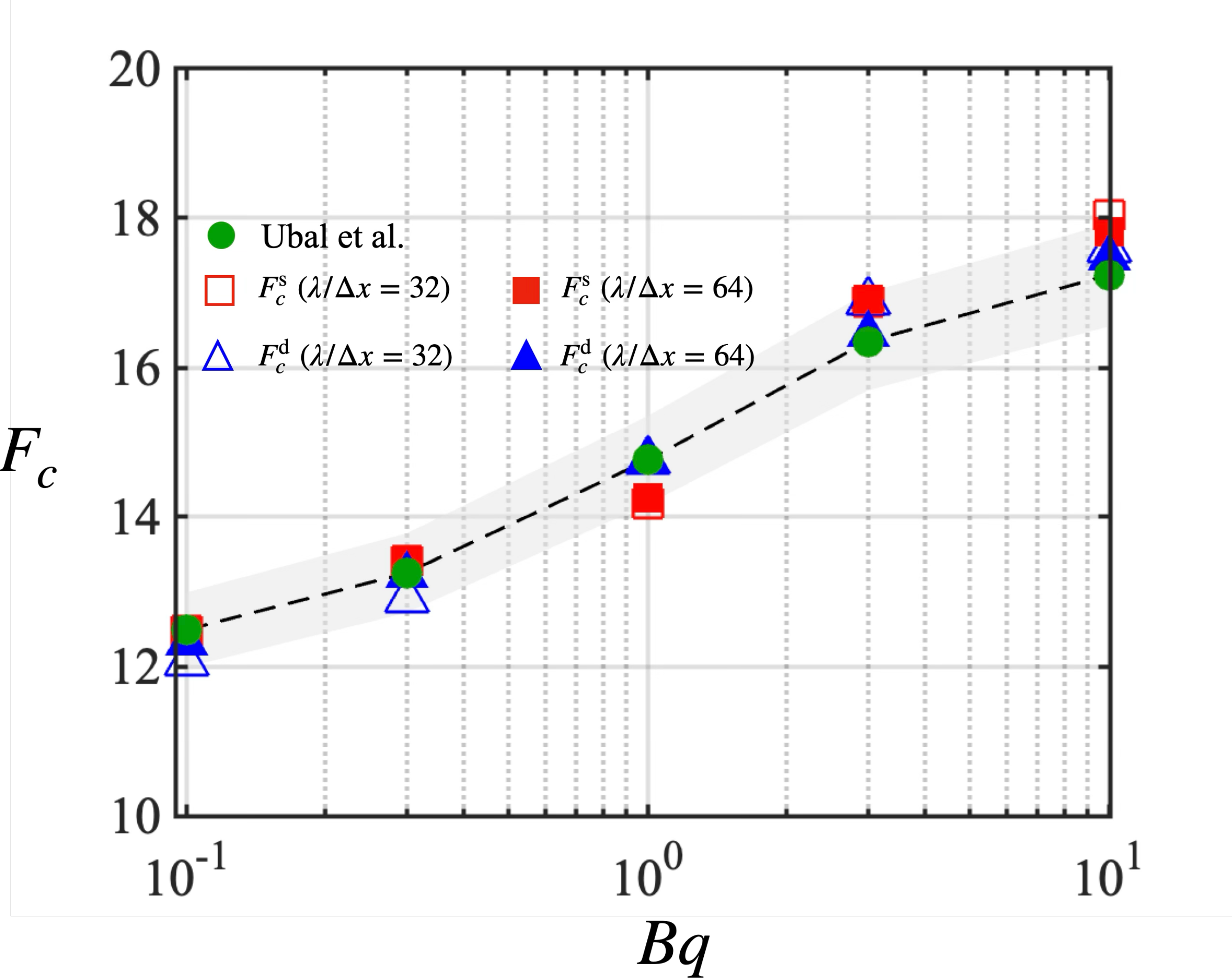}
    \caption{\textbf{Threshold acceleration amplitude evaluation: Threshold acceleration amplitude ($F_c$) is calculated by the growth rates and compared against Ubal et al. (\text{cite}). $F_c^{\rm s}$ and $F_c^{\rm d}$ represent the threshold accelerations evaluated by fixing $Bq_d$ and $Bq_s$ to $0$, respectively. The evaluation is carried out for two different grid sizes, that is, $32$ and $64$ grids across the wavelength. The gray shaded area represents a deviation of maximum $4\%$ from the literature.}}
    \label{fig:far_1}
\end{figure}
We exercised a similar process to vary $Bq_s$ and $Bq_d$ as shown in figure \ref{fig:far_1}. It should be noted that Ubal et al. \cite{ubal2005influence} described a combined Bouossinesq number $Bq = Bq_s + Bq_d$ because the surface viscous stresses reduce to a combined form in a two-dimensional regime. However, the implementation of shear and dilatational surface viscous stresses is not the same. Therefore, we assessed the validity of both $Bq_s$ and $Bq_d$ for this problem. First, we set $Bq_s = 0$ and vary $Bq_d = Bq$ as shown by the triangle markers, and then set $Bq_d = 0$ and vary $Bq_s=Bq$ as shown by the square markers. The filled markers highlight a finer resolution of $\lambda/\Delta x = 64$, whereas the unfilled markers represent the coarser mesh of $\lambda/\Delta x = 32$. It is evident that the threshold acceleration to destabilise a surface-viscous interface increases with an increase in $Bq$. This demonstrates the damping effect of the surface viscous stress at the interface. 
Among $Bq_s$ and $Bq_d$, the error is found to be higher in the case of surface shear viscosity at a maximum of $4\%$. However, all these cases show remarkable agreement with the findings of Ubal et al. \cite{ubal2005influence}.  

\begin{figure}[h!]
    \centering
    \includegraphics[width=1.0\linewidth]{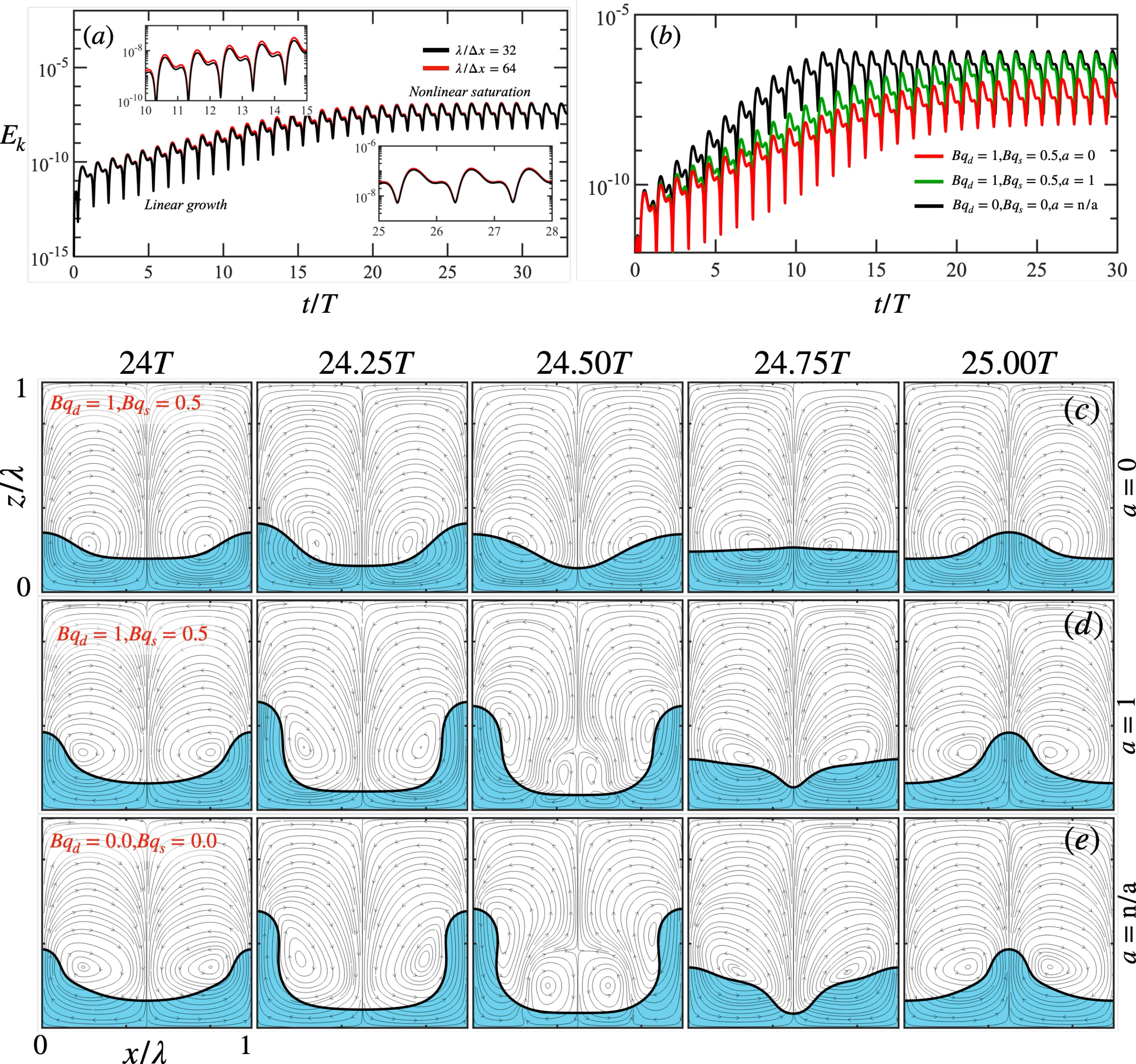 }
    \caption{Temporal evolution of surfactant-covered interface where in panel (a) the interface is surface-viscous of $Bq_d = 1$ and $Bq_s = 0.5$, and $a=0$. In panel (b), the interface is surface viscous with similar properties, but $a=1$. In panels (c-e), the interfacial topology is shown, where blue represents the liquid phase and white represents the gas phase. The interface is highlighted by black contour and is overlaid by the velocity streamlines. $5$ columns in each panel are the snapshots taken at an interval of $0.25T$ from $24T$ to $25T$.}
    \label{fig:far_2}
\end{figure}

Second, we tested the coupled effects of surfactant elasticity and surface viscosity. For this case, we set $Bq_s = 0.5$ and $Bq_d = 1.0$. The surface elasticity is set to $\beta_s = 0.1$, whereas the other dimensionless numbers are kept fixed. At a lower $\beta_s$, the Marangoni stresses due to the surface concentration are so low that the surface acts as a clean surface \cite{panda2024marangoni, pico2024surfactant}. We set the acceleration amplitude $F = 20$, such that it surpassed the threshold acceleration for any of the cases.
In figure \ref{fig:far_2}(a), the temporal evolution of $E_k$ is shown for $33$ time periods for $\lambda/\Delta x = 32$ and $64$. In this case, the surface viscosity is decoupled from the surfactant concentration ($a=0$). Although the grid resolution affects the growth rates and eventually the threshold acceleration calculation, as discussed above and also shown in the inset of figure \ref{fig:far_2}(a) for $t=10T$ to $15T$, the nonlinear saturated state is indifferent for $\lambda/\Delta x = 32$ and $64$ (see the inset of figure \ref{fig:far_2}(a) for $t = 25T$ to $28T$). 
\begin{figure}[h!]
    \centering
    \includegraphics[width=1.0\linewidth]{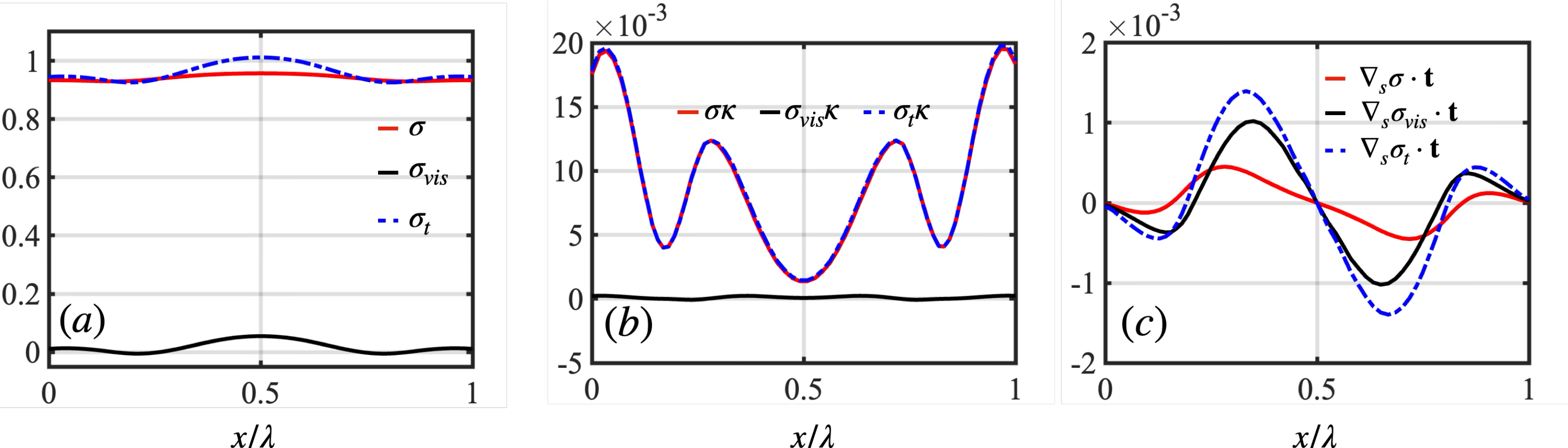}
    \caption{(a) The surfactant-dependent surface ($\sigma$), surface viscous ($\sigma_{vis}$), and total surface ($\sigma_t$) tension along the interface. The normal and tangential stresses due to the surfactant-dependent surface, surface viscosity, and total surface tension are shown in (b) and (c), respectively. }
    \label{fig:far_3}
\end{figure}

In figure \ref{fig:far_2}(b), the temporal evolution of $E_k$ is shown for the cases where $a=0$ and $a=1$ as well as for the case where the surface viscous effects are neglected ($Bq_s = Bq_d = 0$) but $\beta_s = 0.1$. For $a=0$, the surface viscous stresses are maximum; therefore, the damping effects are maximum. Therefore, the kinetic energy increased at the lowest rate. For $a=1$, the surface viscous stresses are a function of the surfactant concentration, and the damping effect reduces significantly whenever the surface is dilated. Thus, surface waves grow faster than in the case $a=0$. When the surface viscous effects were completely ignored, the kinetic energy increased the fastest. Another stark difference is observed in the time of nonlinear saturation of the kinetic energy. The surfactant-covered interface without surface viscous effects shows the quickest saturation of the kinetic energy at $t \approx 10T$, while in the cases of $Bq \neq 0$, the saturation occurred at $t \approx 20T$. At the state of saturation, the surface-viscous interface with $a=0$ shows the least kinetic energy, followed by $a=1$. Interestingly, the kinetic energy for $Bq \neq 0$ but $a=1$ shows almost similar behaviour as that of no surface viscous effects at the nonlinear saturated state. This is further shown in figure \ref{fig:far_2}(c-e), where the interface evolution in the $x-z$ plane at $y = \lambda/4$ is shown at an interface of $0.25T$ and between $24T$ and $25T$. For the case of $Bq_s = 0.5, Bq_d = 1.0$, the viscous effects on the surface dramatically changed the position of the interface, compared to the cases covered with surfactants. For instance, the crest formed at the boundaries of the domain at $t = 24.5T$ shows a stark difference in height for the surface-viscous and surfactant-covered interfaces. Furthermore, a crater-like shape is observed in the surfactant-covered case which is absent in the surface-viscous case at $t=24.75T$. However, when $a=1$, the surface viscous effects are weak, and the interfacial evolution in the nonlinear states is similar to that of the surfactant-covered cases.  


\begin{figure}[h!]
    \centering
    \includegraphics[width= \linewidth]{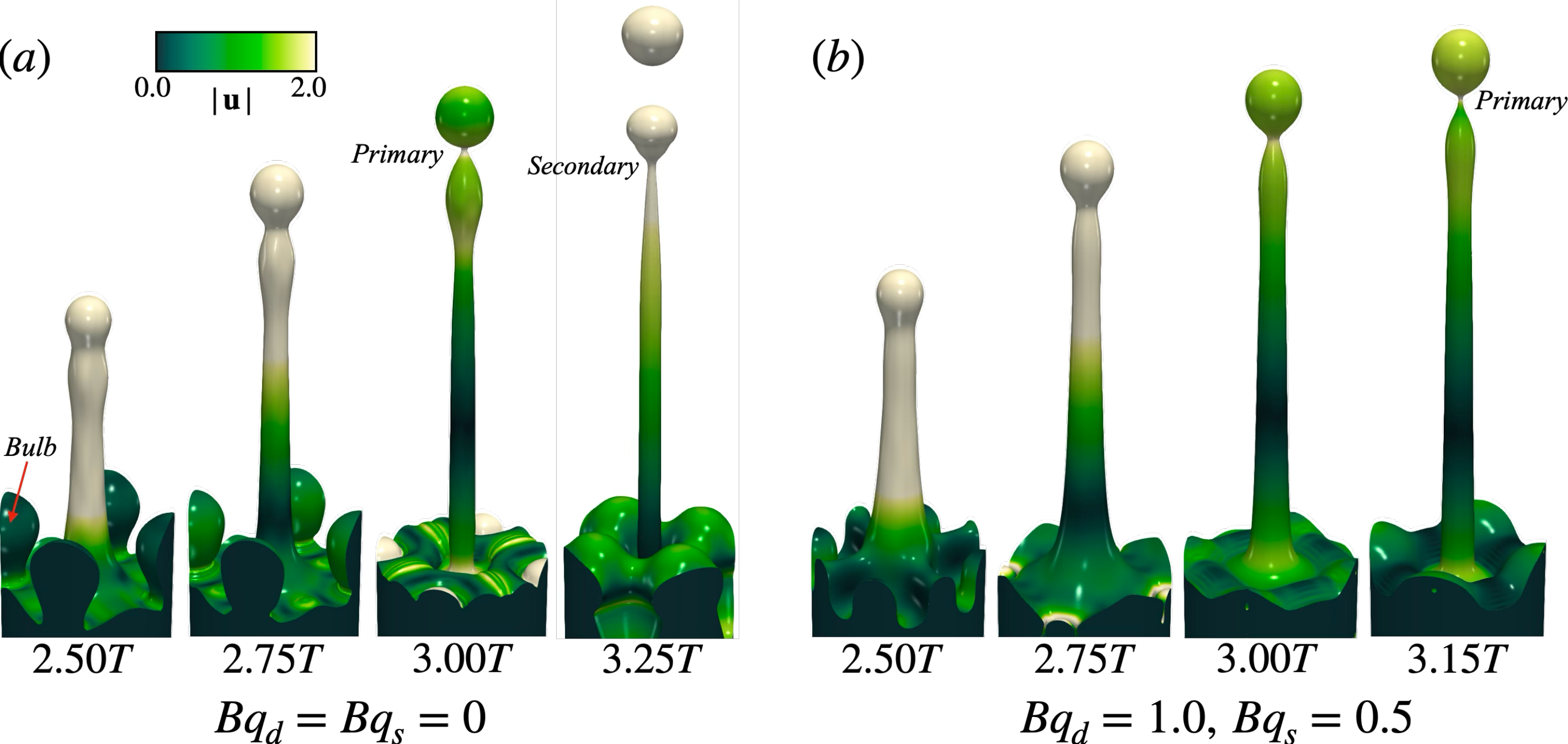}
    \caption{Faraday wave atomisation for a (a) clean and (b) surface viscous contaminated interace: 3D interface is shown at regular intervals and colour-coded by the magnitude of velocity. }
    \label{fig:far_4}
\end{figure}

The surface tension as a function of $\Gamma$, ($\sigma$), surface viscous tension, $\sigma_{vis}$, and total surface tension ($\sigma_t = \sigma + \sigma_{vis}$) is shown in figure \ref{fig:far_3}(a) at $t = 24T$. The variation in surface tension due to the presence of surfactants is negligible, whereas $\sigma_{vis}$ changes significantly along the interface. $\sigma_{vis}$ is the highest at the trough and becomes negative at the points of inflection between the trough and the crest. Although the magnitude of $\sigma_{vis}$ is lower than $\sigma$, the variation in total surface tension $\sigma_t$ is significantly affected by the presence of $\sigma_{vis}$. Figure \ref{fig:far_3}(b) shows that the presence of $\sigma_{vis}$ plays an insignificant role in the normal stress exerted at the interface. However, owing to the significant variation in $\sigma_{vis}$ along the interface, $\nabla_s \sigma_{vis}$ is stronger than $\nabla_s \sigma$ as shown in figure \ref{fig:far_3}(c). The total tangential stress is also significantly affected by the presence of $\sigma_{vis}$. 

Our third test case to demonstrate the relevance of surface viscosity is the atomisation of a surface owing to vibration. In this case, we set the frequency $f = 1000~ \rm Hz$. The density and viscosity ratios are $M_\rho = 10^{-3}$ and $M_\mu = 10^{-2}$, respectively. We set the wavelength, $\lambda = $ $1.219\times 10^{-3} ~ \rm m$, and acceleration amplitude, $F = 688.3$. The Reynolds and Weber numbers are, $Re = 14860$ and $We = 26$, respectively. At such high $Re$, $We$, and $F$, our DNS code is suitable for studying the effects of complex interfaces. The choice of our parameters are aligned with the previous numerical simulations of Faraday wave atomisation \cite{li2020three}.  The three-dimensional visualisation is shown in figure \ref{fig:far_4}(a) for the clean case ($Bq_d = Bq_s = 0$). The troughs are not simple as in the previous case (figure \ref{fig:far_2}). Here, the troughs become extreme craters which eventually burst into a ligament jet that travels upward. Owing to capillary forces, the ligament jet is broken into droplets. As shown in figure \ref{fig:far_4}(a), the ligament jet is broken by pinching off at $t = 3T$. This is the primary pinch-off. Next, at $t = 3.25T$, another pinch-off occurs to obtain a secondary droplet. However, when the interface is surface viscous, that is, $Bq_d = 1, Bq_s = 0.5$, the process of pinching-off of the ligament is derailed. The primary pinch-off occurred at $t = 3.15T$. Moreover, significant changes are observed at the vibrating interface, where a bulk structure is observed in the clean case, but none of such structures are observed in the surface-viscous interface. 
\begin{figure}[h!]
    \centering
    \includegraphics[width= 0.45\linewidth]{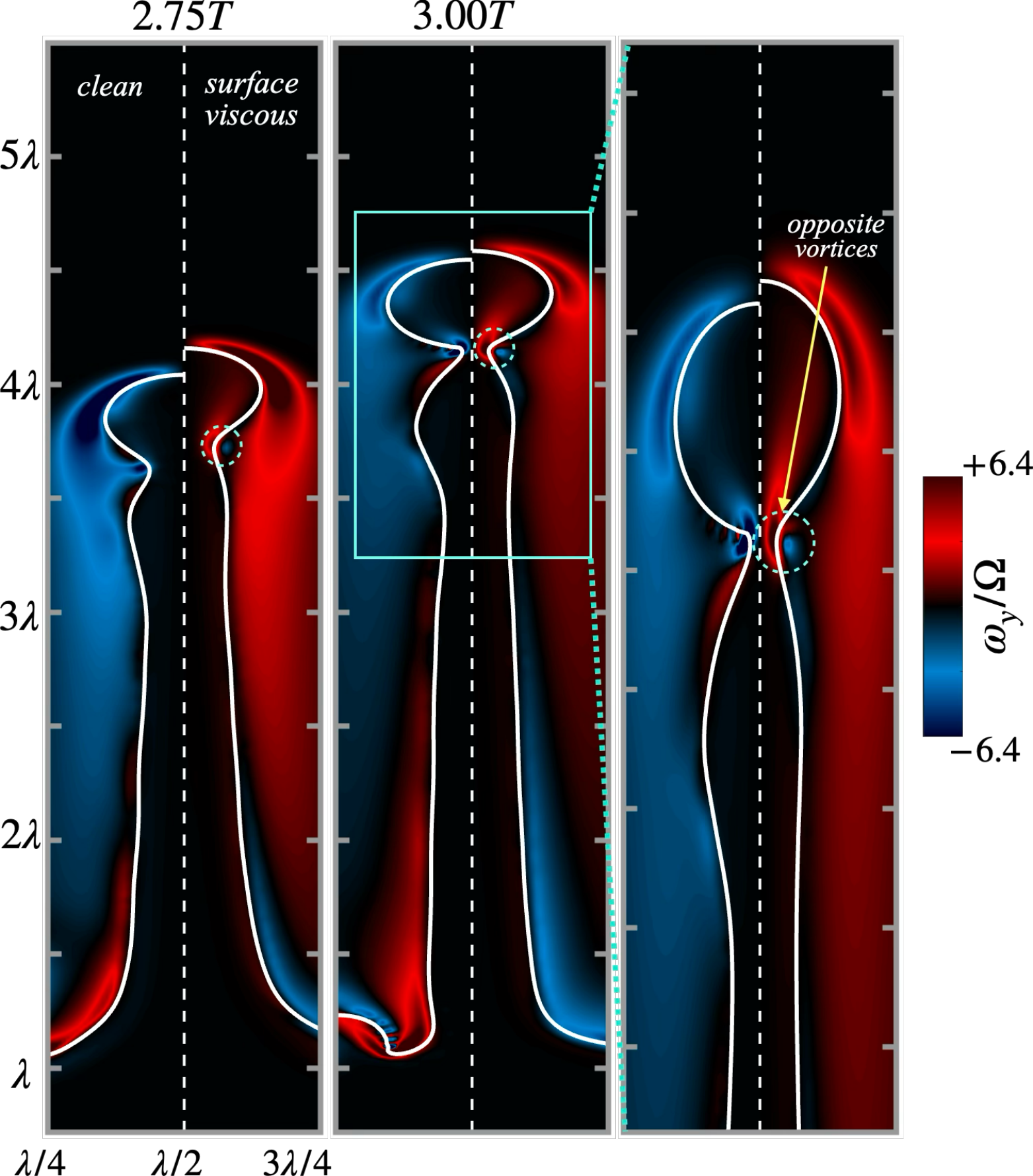}
    \caption{Vorticity contours of the clean (left) and surface viscous contaminated interface at $t = 2.75T$ and $3.00T$. The interface is highlighted by the white line and the cyan box highlight the part of the contour magnified to compare the mechanics of pinch-off at the neck.}
    \label{fig:far_5}
\end{figure}

The delayed pinch-off is due to the tangential stresses that play a role at the neck of the pinching zone. This is shown in figure \ref{fig:far_5}, where opposite vortices are observed near the neck, as opposed to the absence of such flow structures in the clean case. The opposite vortices create a rigidified zone, thus delaying the pinching process \cite{constante2021dynamics}. A more detailed understanding of the physics of atomisation in the presence of surface-viscous surfactants is therefore necessary.    

\section{Conclusion}
A general implementation of 3D interfacial flows with Boussinesq-Scriven surface is proposed in the context of the Level Contour Reconstruction Method (LCRM). The LCRM is based on the implementation of surface tension-driven interfacial flows, leveraging the advantages of front-tracking and level-set methods. Not only is the method robust, but our proposed method is also implemented on massively parallel distributed computing architectures.

We discussed the derivation of the 2D surface viscosity by drawing analogies with the 3D bulk viscosity. A proper description of the 2D rate of deformation tensor is elucidated in various viewpoints. Based on a literature survey, we chose the Lagrangian-extrinsic viewpoint to describe the surface rate of deformation tensor. The Lagrangian-extrinsic viewpoint is the most suitable choice in the context of LCRM. Subsequently, we proposed numerical schemes for evaluating dilatational and shear surface viscous forces. 

To evaluate the dilatational surface viscous forces, we first evaluated a scalar quantity, analogous to the surface tension coefficient: surface viscous tension. Subsequently, we utilised a hybrid surface viscous tension calculation to evaluate the normal and tangential forces owing to the presence of surface viscous tension. One main ingredient in finding the viscous tension of the surface is to construct the tensor of the surface gradient of the interfacial velocity. By interpolating the gradient of the velocity tensor on the Eulerian grid to the midpoint of the edges of each Lagrangian element, we constructed the surface gradient tensor of the interfacial velocity. The trace of this tensor results in the surface divergence of the velocity, and thus the surface viscous tension is evaluated on-the-fly. To evaluate the shear surface viscous forces, the surface gradient tensor of the velocity and its transpose were evaluated at the midpoint of the edges of each Lagrangian element. Using the Gauss divergence theorem, we evaluated the shear surface viscous forces for each element.   

The proposed numerical method is tested for three different cases. The shear surface viscous forces were validated by testing them against the classical case of a neutrally buoyant drop under shear flow. A mesh convergence test and validation against theoretical predictions were performed for $\mu_d = 0$. Then, by fixing $\mu_d=\mu_s$, the implementation of the surface shear forces is validated against the theory and boundary element simulations. To validate the dilatational surface viscous forces, a buoyant drop rising case is tested with $\mu_s = 0$. By increasing the dilatational surface viscosity, the terminal velocity of the rising drop is reduced. Quantitatively, the terminal velocities for varying $\mu_d$ were well validated with the theoretical predictions. Finally, to test the combined effects of both shear and dilatational surface viscous effects, we tested the code for parametric surface waves. We evaluated the threshold acceleration for non-zero dilatational and shear surface viscosities and compared it with previous simulations. A maximum of $4\%$ error is found, in which the error in the cases of non-zero shear surface viscosity is higher than that in the dilatational cases. Then, for non-zero dilatational and shear surface viscosities, nonlinear surface waves were tested for surfactant-dependent cases. Finally, we presented the utility of our code for impulsive surface waves where atomisation occurs. The presence of surface viscous effects can cause the damping of atomised jets, leading to a delay in drop formation. However, an in-depth study of the underlying physics is beyond the scope of this study and will be addressed in future work. 

\section*{Acknowledgements}
This work was supported by the Engineering and Physical Sciences Research Council, UK, through the  PREMIERE (EP/T000414/1) programme grant, the ANTENNA Prosperity Partnership grant (EP/V056891/1), and by the National Research Foundation of Korea(NRF) grant funded by the Korea government (MSIT) (No. RS-2025-02302984). D.P., A.M.A., and L.K. acknowledge HPC facilities provided by the Imperial College London Research Computing Service.  
D.P. and L.K. acknowledge Dr. Paula Pico for her efforts to introduce surface viscosity during her doctoral studies. 
D.P. acknowledges the Imperial College London President’s PhD scholarship. A.M.A. acknowledges the Kuwait Foundation for the Advancement of Sciences (KFAS) for his financial support. D.J. and J.C. acknowledge support through HPC/AI computing time at the Institut du Developpement et des Ressources en Informatique Scientifique (IDRIS) of the Centre National de la Recherche Scientifique (CNRS), coordinated by GENCI (Grand Equipement National de Calcul Intensif) 
grant 2025 A0182B06721. The numerical simulations were performed using the BLUE code \cite{shin2017solver} and the visualisations were generated using ParaView.

\appendix 
\section{Interpretation of surface deformation rate tensor}\label{appA}
Lopez and Hirsa \cite{lopez1998direct} interpreted the surface deformation rate tensor as the symmetric part of the surface gradient ($\nabla_s = \mathbf I_s \nabla$) of the surface velocity ($\mathbf u_s = \mathbf I_s \mathbf u$) vector and projected on the surface by the surface projection tensor $\mathbf I_s$. However, there is a bottleneck in such formulations for generalised flows. In this way, we write the surface gradient of the surface velocity as,
\begin{equation}\label{lopez1}
    \nabla_s \mathbf u_s = (\mathbf I_s \cdot \nabla)(\mathbf I_s \cdot \mathbf u_f) = \mathbf I_s \cdot (\nabla (\mathbf I_s \cdot \mathbf u_f)).
\end{equation}
Applying the gradient of the surface velocity, we find, 
\begin{equation}\label{lopez2}
    \mathbf I_s \cdot (\nabla (\mathbf I_s \cdot \mathbf u_f))= \mathbf I_s \cdot (\nabla \mathbf I_s \cdot \mathbf u_f + \mathbf I_s \cdot \nabla \mathbf u_f).
\end{equation}
Since, $\mathbf I_s$ is symmetric and idempotent, \eqref{lopez2} is written as,
\begin{equation}\label{lopez3}
    \nabla_s \mathbf u_s = \mathbf I_s \cdot \nabla \mathbf I_s \cdot \mathbf u_f + \mathbf I_s \cdot \nabla \mathbf u_f \implies (\nabla_s \mathbf I_s)\cdot \mathbf u_f + \mathbf I_s \cdot \nabla \mathbf u_f 
\end{equation}
Surface projection of equation \eqref{lopez3} gives,
\begin{equation}\label{lopez4}
    \nabla_s \mathbf u_s \cdot \mathbf I_s = (\nabla_s \mathbf I_s)\cdot \mathbf u_f \cdot \mathbf I_s + \mathbf I_s \cdot \nabla \mathbf u_f \cdot \mathbf I_s
\end{equation}
The second term in Equation \eqref{lopez4} is the surface projection of the deformation rate tensor, as obtained by Secomb and Kalak \cite{secomb1982surface}. The first term is the contraction of a third-rank tensor $\nabla_s \mathbf I_s$ with the vector $\mathbf u_f$. Deriving $\nabla_s \mathbf I_s$ further, we obtain,
\begin{equation}\label{lopez5}
    \nabla_s \mathbf I_s = \mathbf I_s \cdot \nabla \mathbf I_s \cdot \mathbf I_s\implies \mathbf I_s \cdot (\nabla \mathbf I - \nabla (\mathbf n \otimes \mathbf n))\cdot \mathbf I_s.
\end{equation}
Since, gradient of an identitiy tensor is a null tensor and $\mathbf n \cdot \mathbf I_s = 0$, \eqref{lopez5} is written as, 
\begin{equation}
    \nabla_s \mathbf I_s  = -\mathbf I_s \cdot (\mathbf K \otimes \mathbf n + \mathbf n \otimes \mathbf K)\cdot \mathbf I_s,
\end{equation}
where $\mathbf K = \nabla \mathbf n$ is the curvature tensor. This results in the surface projection of a third-rank tensor which is a product of the curvature tensor, $\mathbf K$ and the normal vector, $\mathbf n$. In the cases of linear analyses and other approximations, such as two-dimensional flows and lubrication theory, such a high-ranked tensor implicitly drops to a null tensor. Moreover, if we specifically analyse a flat interface, as studied by \cite{lopez1998direct}, the curvature of the surface $\approx 0$ which makes the curvature tensor negligible. However, in generalised flows, the curvature tensor may not be $0$ therefore, the surface gradient of the surface velocity is not an appropriate choice for the surface deformation rate tensor.

\bibliographystyle{elsarticle-num}
\bibliography{main_biblio}
\end{document}